\documentclass[aps,pre,reprint,showpacs,twocolumn,nofootinbib]{revtex4}
\usepackage{amsmath,amssymb,bbm,hyperref,graphicx,color}

\begin{document}

\title{Non-equilibrium Relaxation and Aging Scaling of the Coulomb and Bose
       Glass}

\author{Matthew T. Shimer$^{1,2}$}
\author{Uwe C. T\"auber$^{1}$} 
\author{Michel Pleimling$^{1}$}

\affiliation{$^1$Department of Physics (MC 0435), Robeson Hall, 
  850 West Campus Drive, Virginia Tech, Blacksburg, VA 24061}
\affiliation{$^2$hMetrix LLC, 150 Monument Rd, \# 107, Bala Cynwyd, PA 19004}

\date{\today} 


\begin{abstract}
We employ Monte Carlo simulations to investigate the non-equilibrium 
relaxation properties of the two- and three-dimensional Coulomb glass with
different long-range repulsive interactions.
Specifically, we explore the aging scaling laws in the two-time density 
autocorrelation function.
We find that in the time window and parameter range accessible to us, the 
scaling exponents are not universal, depending on the filling fraction and
temperature:
As either the temperature decreases or the filling fraction deviates more 
from half-filling, the exponents reflect markedly slower relaxation kinetics. 
In comparison with a repulsive Coulomb potential, appropriate for impurity 
states in strongly disordered semiconductors, we observe that for logarithmic
interactions, the soft pseudo-gap in the density of states is considerably 
broader, and the dependence of the scaling exponents on external parameters
is much weaker.
The latter situation is relevant for flux creep in the disorder-dominated 
Bose glass phase of type-II superconductors subject to columnar pinning
centers.
\end{abstract}

\maketitle

\section{Introduction} 
\label{intro}

The Coulomb glass model was devised to describe the physical properties of
localized charge carriers in disordered semiconductors 
\cite{Efros75, Shklovskii84, Efros85}.
It assumes that the localization length $\xi$ is small or of the order of
the mean separation $a_0$ between acceptor or donor sites, whence the system
can be essentially described in classical terms: 
Charged particles are confined to randomly distributed sites, and at low
temperatures the system equilibrates through rearrangement of the carrier
distribution to minimize the total interaction energy.
In semiconductors, these variable-range hopping processes are effected
through phonon-assisted tunneling between the acceptor / donor sites. 
The strong spatial (anti-)correlations resulting from the long-range 
repulsive forces in turn induce a marked depletion of the (interacting) 
single-particle density of states, i.e., the  distribution function 
$g(\epsilon)$ of the site energies, near the chemical potential $\mu_c$ that
separates low-energy ($\epsilon < \mu_c$) filled states from empty states at
elevated energies ($\epsilon > \mu_c$).
In the presence of this correlation-induced soft Coulomb gap, carrier
mobility thus becomes considerably impeded \cite{Shklovskii84}:
If $g(\epsilon) \sim |\epsilon - \mu_c|^\gamma$ follows a power law in the
vicinity of $\mu_c$ with an (effective) gap exponent $\gamma$, the associated
conductivity scales as $\ln \sigma \sim - T^{- p}$, with 
$p(\gamma) = (\gamma + 1) / (\gamma + d + 1)$ in $d$ spatial dimensions in the
thermally activated transport regime at low temperatures $T$.
Note that $p(\gamma) \geq 1 / (d + 1) = p(\gamma \to 0)$, the Mott 
variable-range hopping exponent applicable for a finite density of states 
$g(\mu_c) > 0$.
Electron tunneling experiments in doped semiconductors have confirmed the 
existence of correlation-induced soft gaps in the density of states
\cite{Massey95, Butko00}.

The two-dimensional Coulomb glass model, with the electrostatic $1/r$ 
potential essentially replaced by a logarithmic repulsion, has furthermore 
been adapted to capture the static properties as well as thermally activated
flux creep in type-II superconductors with extended, linear disorder aligned
along the magnetic-field direction \cite{Blatter94}.
These columnar defects serve as effective pinning sites for fluctuating 
magnetic flux lines; at low temperatures $T$ (and driving currents $J$) they
undergo a continuous localization transition \cite{Lyuksyutov92, Nelson92}.
In this localized Bose glass phase, the pinned flux lines are essentially
straight and parallel, rendering the system effectively two-dimensional, and 
vortex transport between columnar defects proceeds in analogy to 
variable-range hopping through formation and subsequent relaxation of double 
kinks between different pinning sites \cite{Nelson93, Dai95}.
Long-range repulsive vortex interactions again induce a soft gap in the
density of states which strongly suppresses flux creep, leading to a
desired much reduced resistivity $\ln \rho \sim - J^{- p} / T$ 
\cite{Tauber95} (for magnetic flux densities smaller than the matching field,
at which the number of flux lines equals the number of columnar defects; for
extensions to the regime near and beyond the matching field, see 
Refs.~\cite{Wengel97, Wengel98}).

Over the past three decades, intense research into the correlation-dominated 
equilibrium features \cite{Davies82, Grunewald82, Davies84, Levin87, Xue88, 
Mobius92, Grannan93, Menashe00, Menashe01, Muller07, Surer09, Goethe09, 
Efros11} as well as non-equilibrium relaxation properties \cite{Perez99, 
Yu99, Diaz01, Tsigankov03, Grempel04, Kolton05, Amir08, Amir09, Kirkengen09, 
Oreg09, Shimer10, Amir11, Borini11, Meroz14, Amir14} of the Coulomb glass 
have considerably advanced our understanding of this paradigmatic model 
system for highly correlated disordered materials. 
In part motivated by the unambiguous experimental confirmation of aging 
effects in relaxation measurements for the conductivity of a two-dimensional
silicon sample, and scaling near its metal-insulator transition 
\cite{Jaroszynski07, Popovic07}, in this work we focus on a numerical study
of the non-equilibrium relaxation properties of the Coulomb glass following
a quench from a fully uncorrelated, high-temperature initial state.

Although the soft Coulomb gap in the site energy distribution forms quite
fast, subsequent relaxation towards the equilibrium terminal state is
sufficiently slow to open a sufficiently wide time window wherein time
translation invariance is broken and aging scaling is clearly observed (for
recent overviews on non-equilibrium relaxation and aging phenomena, see 
Refs.~\cite{Henkel07, Henkel09}).
Specifically, we employ a variant of the Monte Carlo algorithm described in
Refs.~\cite{Grempel04, Kolton05} to investigate the dependence of the ensuing 
aging scaling exponents for various Coulomb glass systems as function of 
temperature $T$, filling fraction (total charge carrier density) $K$, 
dimensionality, and form of the repulsive interaction potential (Coulomb 
$1 / r$ potential in $d = 2, 3$ dimensions; logarithmic potential in two 
dimensions).
A first, concise account of aging in the two-dimensional Coulomb case was
presented in Ref.~\cite{Shimer10}; we note that further technical details
and additional data can be found in Ref.~\cite{Shimer11}.

In the following Sec.~\ref{model}, we introduce our model Hamiltonian and
explain our Monte Carlo simulation algorithm.
We also briefly discuss basic simulation results pertaining to the emerging
Coulomb gap in the (interacting) single-particle density of states.
Section~\ref{aging} addresses non-equilibrium relaxation properties of our
system as obtained from measurements of the two-time density autocorrelation
function, starting from random initial conditions.
The obtained aging scaling exponents and their dependence on temperature,
filling fraction, interaction potential, and dimensionality constitute the
central findings of this work.
We conclude with a brief summary and discussion.

\section{Model Description and Monte Carlo Simulations}
\label{model}

In this section, we briefly describe the Coulomb glass model, explain our
Monte Carlo algorithm, and list our results on equilibrium properties 
obtained from our simulation runs in two and three dimensions with different
interaction potentials.

\subsection{The Coulomb Glass Model}

The Coulomb glass model was introduced by Efros and Shklovskii to capture 
thermodynamic and transport properties of localized charge carriers in doped
semiconductors \cite{Efros75}. 
A set of multiple, randomly (Poisson) distributed but fixed localized pinning 
sites (here selected off-lattice on a continuum) are available to the charge 
carriers in $d$ spatial dimensions. 
Because of the strong intra-site correlations these sites labeled by an index
$i$ can only contain at most a single particle, which restricts the site 
occupation numbers to $n_i = 0, 1$. 
The system is dominated by long-range repulsive interactions $V(r)$ between
the charge carriers.
The combination of quenched spatial site disorder and long-range interactions
induce strong correlation effects.

For the case of unscreened Coulomb interactions, the Hamiltonian of the 
Coulomb glass model reads \cite{Efros75, Shklovskii84}
\begin{equation}
\label{cgla}
  H(\{ n_i \}) = \sum_i n_i \varphi_i + \frac{e^2}{2 \kappa} \sum_{i \neq j} 
      \frac{(n_i - K) (n_j - K)}{|{\bf R}_i - {\bf R}_j|} \, ,
\end{equation}
where $e$ denotes the carrier charge, $\kappa$ a dielectric constant, and
${\bf R}_i$, $\varphi_i$, and $n_i$ respectively represent the position 
vector, (bare) site energy, and occupancy of the $i$th site, 
$i = 1, \ldots, N$. 
The first term corresponds to (random) site energies assigned to each 
accessible location; since the system is dominated by the long-range forces, 
we choose all $\varphi_i = 0$ to further simplify the model, while drawing 
the positions ${\bf R}_i$ at random from a two- or three-dimensional 
continuous set \cite{Xue88, Yu99, Grempel04, Kolton05}. 
The second contribution encapsulates the repulsive Coulomb interactions (with
dielectric constant $\kappa$). 
In order to maintain global charge neutrality, a uniform relative charge 
density $K = \sum_i n_i / N$ is inserted; it constitutes the total carrier 
density per site or filling fraction. 
Note that with $\varphi_i = 0$ the Hamiltonian (\ref{cgla}) displays 
particle-hole symmetry, i.e., systems with filling fractions $K = 0.5 + k$ 
and $K = 0.5 - k$ are equivalent.
Upon replacing the site occupation numbers with Ising spin variables 
$\sigma_i = 2 n_i - 1 = \mp 1$, the Coulomb glass maps onto a random-site, 
random-field antiferromagnetic Ising model with long-range exchange
interactions \cite{Davies82}.

The Coulomb glass model may be adapted to describe the low-temperature 
properties of magnetic flux lines in type-II superconductors with strong 
columnar pinning centers \cite{Nelson93, Dai95, Tauber95}.
Deep in the Bose glass phase, the vortices become localized at the linear
material defects, and thermal transverse wandering is strongly suppressed,
which renders the system essentially two-dimensional.
The mutual repulsion interaction between two occupied sites is now 
characterized by a modified Bessel function $K_0(r / \lambda)$, essentially
a long-range logarithmic potential that is screened on the scale of the 
London penetration depth $\lambda$, and the Hamiltonian becomes
\begin{equation}
\label{bgla}
  H(\{ n_i \}) = \epsilon_0 \sum_{i \neq j} (n_i - K) (n_j - K) \, 
  K_0\!\left( \frac{|{\bf R}_i - {\bf R}_j|}{\lambda} \right) .
\end{equation}
The energy scale is now set by $\epsilon_0 = (\phi_0 / 4 \pi \lambda)^2$ 
with the magnetic flux quantum $\phi_0 = h c / 2 e$.
We shall address the dilute low-magnetic field regime where all 
site distances $r_{ij} =  |{\bf R}_i - {\bf R}_j| \ll \lambda$ and thus
$K_0(x) \approx - \ln x$ (aside from a constant).
The random site positions ${\bf R}_i$ are drawn from a continuous flat 
distribution in two dimensions.

\subsection{Monte Carlo Simulation Algorithm}

The Monte Carlo simulations were initiated by randomly placing $N$ sites 
within a square in two / cube in three 
dimensions. 
Initially, we prepared the system in a completely uncorrelated configuration,
distributing $K N$ charge carriers at random among the $N$ available sites. 
The ``charged'' particles may then attempt hops from occupied sites $a$ 
(with $n_a = 1$) to unoccupied sites $b$ ($n_b = 0$). 
Following Refs.~\cite{Grempel04, Kolton05}, two multiplicative factors 
determine the success rate of this hop, namely (i) a strongly 
distance-dependent transfer process that respectively models 
phonon-mediated tunneling in semiconductors, and vortex superkink 
proliferation in type-II superconductors, and (ii) thermally activated jumps
over energy barriers represented by a Metropolis factor:
\begin{equation}
\label{rate}
  \Gamma_{a \to b} = \tau_0^{-1} \, e^{- 2 r_{ab} / \xi} \,
  \min[1,e^{- \Delta E_{ab} / T}] \, ,
\end{equation}
where $\tau_0$ represents a microscopic time scale, 
$r_{ij} = |{\bf R}_i - {\bf R}_j|$ is the distance between sites $i$ and 
$j$, while $\xi$ characterizes the spatial extension of the localized 
carrier wave functions / thermal wandering of the magnetic flux lines (we 
set Boltzmann's constant $k_{\rm B} = 1)$. 
The rate for a thermally activated move from occupied site $a$ to empty 
site $b$ is determined by the energy difference 
$\Delta E_{ab} = \epsilon_b - \epsilon_a - V(r_{ab})$, with the 
(interacting) site energies 
$\epsilon_i = \sum_{j \neq i} (n_j - K) \, V(r_{ij})$, and where the 
long-ranged interactions are governed by the Coulomb potential 
$V(r) = e^2 / \kappa r$ for semiconductor charge carriers, whereas 
$V(r) = 2 \epsilon_0 \, K_0(r / \lambda)$ for magnetic vortices.

The simulation consecutively performs the following four stochastic
processes \cite{Shimer10, Shimer11}:
(i) Randomly select an occupied site $a$ ($n_a = 1$). 
(ii) Choose an unoccupied site $b$ ($n_b = 0$) from the exponential 
    probability distribution in the first, ``tunneling'' term in 
    Eq.~(\ref{rate}).
(iii) Attempt a hop with a success probability determined by the Metropolis 
    factor in Eq.~(\ref{rate}). 
(iv) If the hop attempt fails, return to step one.
    If it is successful, move the particle from site $a$ to site $b$.
Each Monte Carlo time step (MCS) consists of $N$ iterations of (i)--(iv).
Note that all pair potential values $V(r_{ab})$ may be calculated at the 
beginning of the simulation run. 
Subsequently, only the site energies of sites $a$ and $b$ need to be 
evaluated, which merely requires a summation of the pre-calculated pair 
potentials. 
Collecting these interacting site energies and averaging over many 
independent realizations with different random site placements, we then 
compiled the (interacting) single-particle density of states $g(\epsilon)$, 
to be discussed in the following subsection. 
With all occupation numbers $n_i$ recorded at each time step, we could
furthermore study the temporal evolution of the two-time carrier density 
autocorrelation function, see Sec.~\ref{aging}.

In the following, distances are measured relative to the mean separation 
$a_0$ between sites, and energies as well as temperature scales are given
in units of the typical energy scales $e^2 / \kappa a_0$ and 
$2 \epsilon_0 \, K_0(a_0 / \lambda)$, where we used $\lambda / a_0 = 8$. 
As in Refs.~\cite{Grempel04, Kolton05}, we set $\xi = a_0$; we have in
fact explored other values for $\xi$ as well, $0.5 a_0$ and $2 a_0$, but
(within the applicability range of the model) found that the ensuing changes
can simply be absorbed into a renormalized overall time scale $\tau_0$.
Initially, $K N$ particles were placed at random on the $N = L^d$ 
available sites to mimic a quench from a very high temperature. 
Then the system was evolved for typically $10^6$ MCS at temperature $T$ with
the Monte Carlo dynamics defined by the generalized Metropolis rate 
(\ref{rate}).
We employed periodic boundary conditions, whence the potential due to 
charges outside the simulation cell was calculated by mirroring it on the 
$2 d$ adjacent faces.
The minimum of the distances between any given sites $i$ and $j$ and
the latter's $2 d$ mirror images in neighboring cells is used to compute
the interaction potential $V(r_{ij})$.

We performed simulations for different system sizes $8 \leq L \leq 32$; with
temperatures in the range $0.001 \leq T \leq 0.1$; and filling fractions in 
the interval $0.25 \leq K \leq 0.5$ (equivalent to $0.5 \leq K \leq 0.75$ 
due to particle-hole symmetry). 
Running the simulations with various system sizes $L$, we noticed no 
measurable finite-size effects; for example, deviations between the obtained
density autocorrelations at $L = 10$ and $L = 16$ were less than $2 \%$ 
\cite{Shimer11}.
For each configuration (temperature $T$, filling fraction $K$, etc.), the
data were averaged over at least $1000$ independent simulation runs. 
Temperatures larger than $0.03$ turned out not to be useful for our study of
aging processes since equilibrium was then reached far too quickly. 
In contrast, for $T < 0.01$, the kinetics slowed down too much for gathering
statistically significant data within computationally reasonable time 
frames. 
As will be discussed in more detail below, the dynamics also freezes out 
within the numerically accessible simulation times for filling fractions
$K < 0.4$ (or $K > 0.6$).

\subsection{Coulomb Gap Properties}

The long-range interactions quickly generate strong correlations among the 
``charged'' particles.
As they maximize their distances subject to the availability of randomly
placed pinning sites, in equilibrium a pronounced soft Coulomb gap forms 
at zero temperature in the (interacting) single-particle density of states, 
or distribution of site energies $g(\epsilon)$ \cite{Shklovskii84}.
Following Efros and Shklovskii's insightful mean-field argument, this
interacting density of states vanishes precisely at the chemical potential 
$\mu_c$ that separates the low-energy filled sites from the more energetic 
empty sites, $g(\mu_c) = 0$.
For a power-law repulsive interaction potential $V(r) \sim r^{- \sigma}$, 
the mean-field analysis further predicts that for $\sigma < d$
\begin{equation}
\label{cgap}
  g(\epsilon) \sim |\epsilon - \mu_c|^\gamma
\end{equation}
vanishes algebraically near $\mu_c$ in $d$ dimensions, with the positive
gap exponent $\gamma = (d / \sigma) - 1$ \cite{Shklovskii84, Tauber95}.
Beyond mean-field theory, this expression still represents a lower bound for
$\gamma$ \cite{Efros11}. 
Indeed, Monte Carlo simulations typically yield gap exponent values that 
exceed the mean-field estimate \cite{Davies82, Davies84, Levin87, Mobius92, 
Tauber95}, especially in the absence of random on-site disorder 
($\varphi_i = 0$).
For example, in their very detailed numerical Coulomb glass study with up to
$N = 125,000$ and $40,000$ sites in $d = 3$ and $d = 2$ dimensions, M\"obius 
and Richter measured $\gamma = 2.6 \pm 0.2$ and $\gamma = 1.2 \pm 0.1$, 
respectively \cite{Mobius92}.
More recent studies, however, found gap exponents much closer to the 
mean-field predictions \cite{Surer09, Goethe09}.
Tunneling experiments on the non-metallic semiconductor Si:B samples yielded
a gap exponent $\gamma \approx 2.2$ \cite{Massey95}, while transport 
measurements on ultrathin Be films were compatible with the mean-field value 
$p = 1/2$, i.e., $\gamma = 1$ for $d = 2$ \cite{Butko00}.
For the two-dimensional Bose glass with essentially logarithmic repulsion, 
the mean-field argument predicts an exponential gap ($\sigma \to 0$); in 
contrast, the data in Ref.~\cite{Tauber95} from zero-temperature simulations
with $N = 400$ sites could best be fitted with power laws and (perhaps just 
effective) gap exponents that increase with decreasing filling fraction $K$, 
ranging from $\gamma \approx 2.2$ for $K = 0.4$ to $\gamma \approx 2.9$ for 
$K = 0.1$.
Indeed, correlation effects should be strongest far away from half-filling, 
since the charged particles are then least affected by the Poissonian 
spatial disorder.

\begin{figure}
(a) \includegraphics[angle=0,width=0.93\columnwidth]{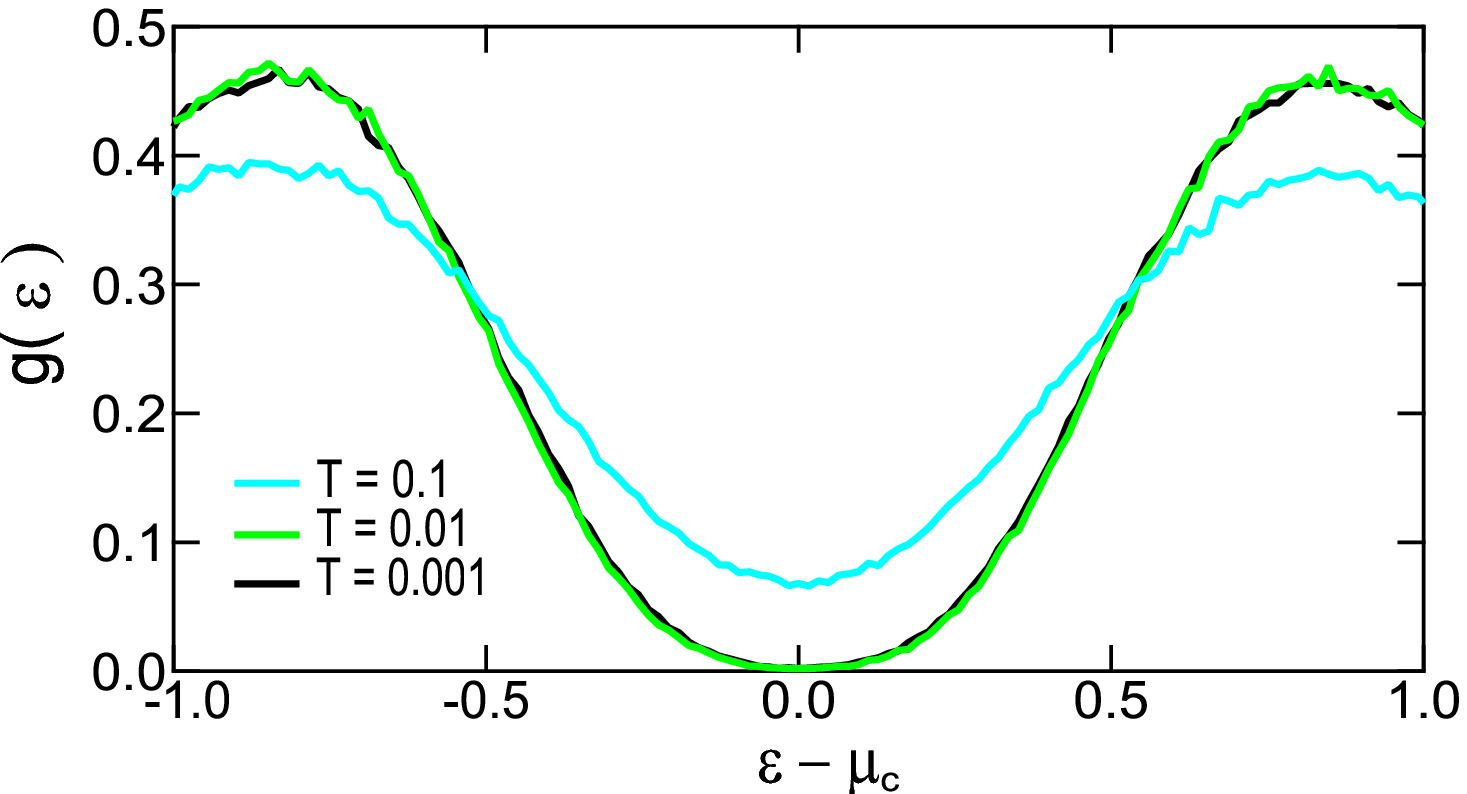} 
\vskip 0.1cm
(b) \includegraphics[angle=0,width=0.93\columnwidth]{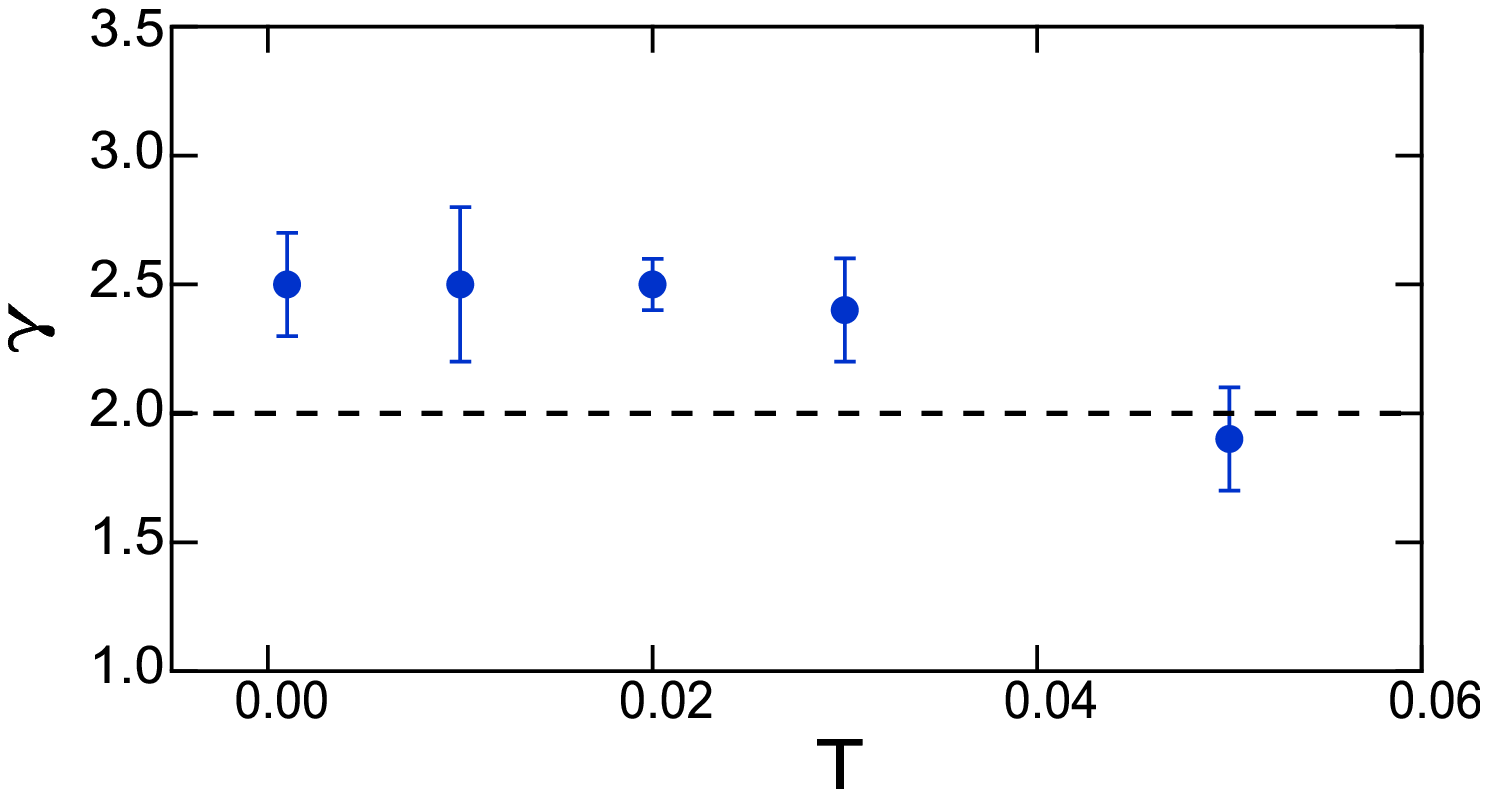}
\vskip 0.1cm  
(c) \includegraphics[angle=0,width=0.93\columnwidth]{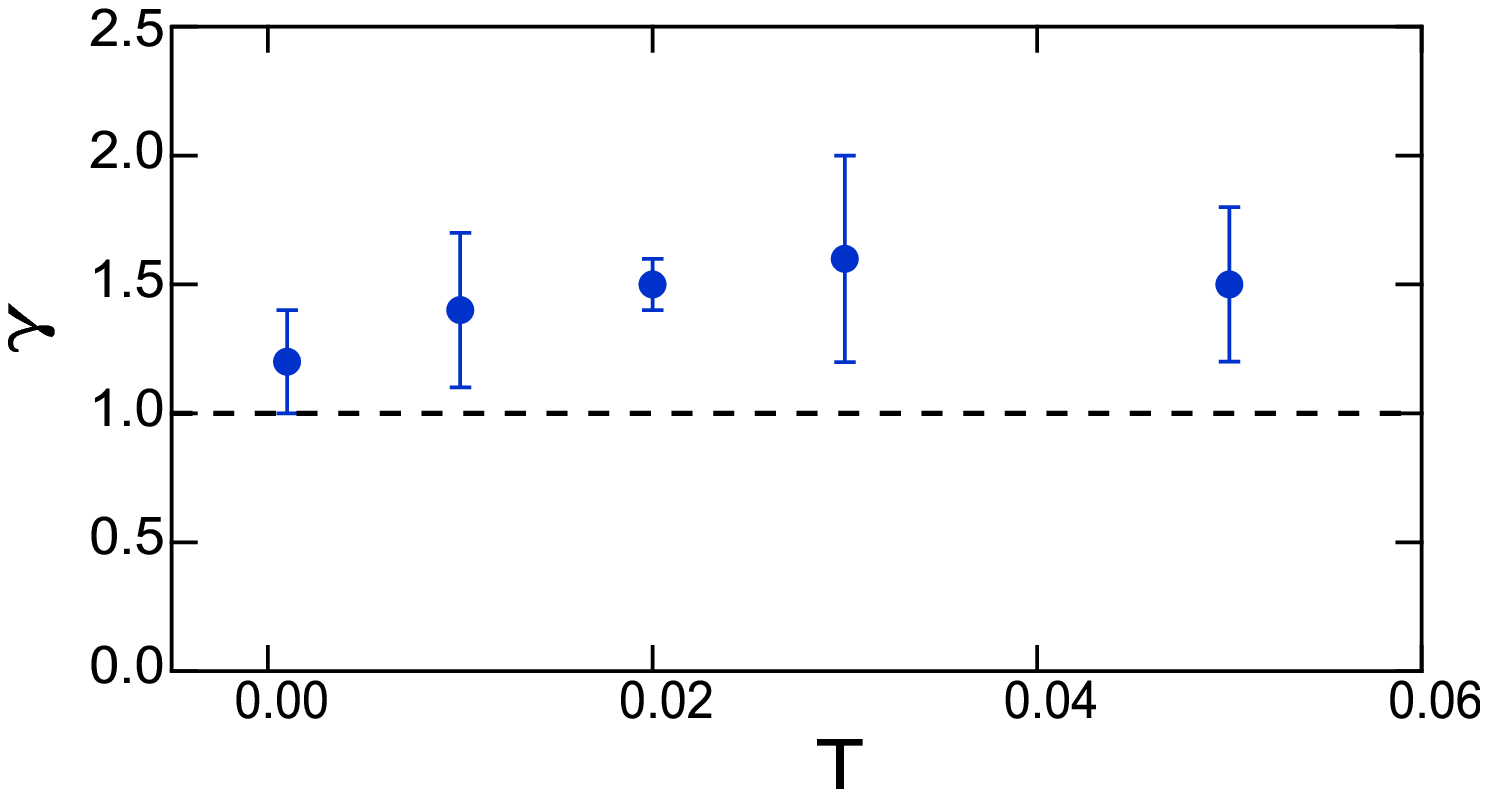}
\caption{(Color online) (a) Density of states $g(\epsilon)$ for the 
   three-dimensional Coulomb glass ($L = 8$, $N = 512$ sites) at 
   half-filling $K = 0.5$, at temperatures $T = 0.001$ (black),
   $T = 0.01$ (green/gray), and $T = 0.1$ (light blue/light gray).
   (b) Corresponding (effective) gap exponents $\gamma(T)$ vs. temperature;
   (c) gap exponent $\gamma(T)$ in two dimensions ($L = 16$, $N = 256$, 
   $K = 0.5$). 
   The dashed lines represent the mean-field prediction $\gamma = d - 1$.} 
\label{fig1}
\end{figure}
In our Monte Carlo simulations performed in the absence of a background
random site energy distribution, the Coulomb gap in the single-particle 
density of states forms very quickly, and appears fully formed within 
$\sim 50 \ldots 100$ MCS \cite{Shimer10, Shimer11}.
To make contact with previous work, we have measured the interacting density
of states / distribution of site energies, and obtained approximate values
for the effective gap exponents $\gamma$ from best linear fits near the 
chemical potential $\mu_c$ in double-logarithmic plots.
In Fig.~\ref{fig1}(a), we display results for the temperature dependence of
the shape of $g(\epsilon)$ for the three-dimensional Coulomb glass with
repulsive $1 / r$ interaction potential at half-filling.
The graphs for $T = 0.001$ and $T = 0.01$ are indistinguishable within the
statistical errors.
\begin{figure}
(a) \includegraphics[angle=0,width=0.93\columnwidth]{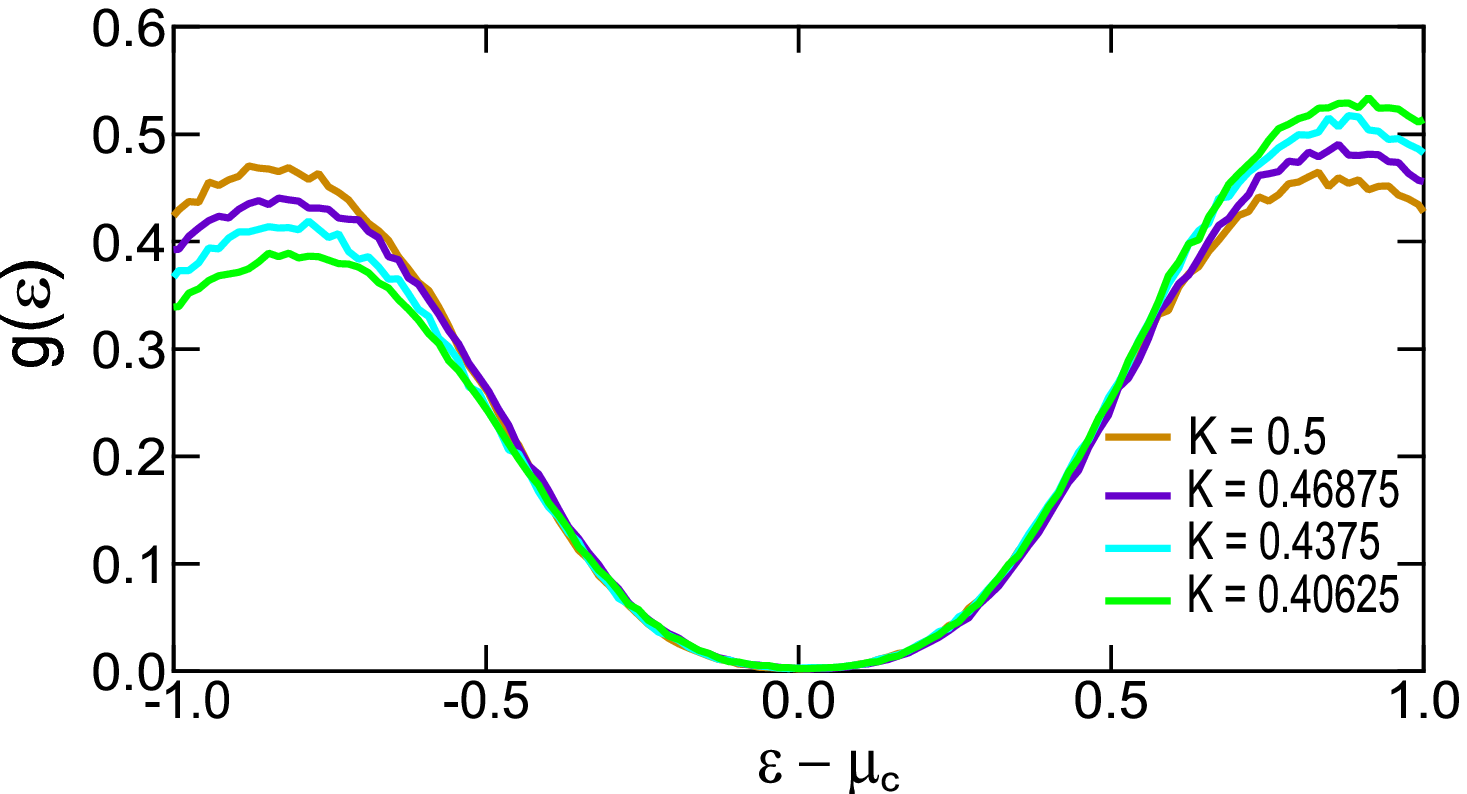} 
\vskip 0.1cm
(b) \includegraphics[angle=0,width=0.93\columnwidth]{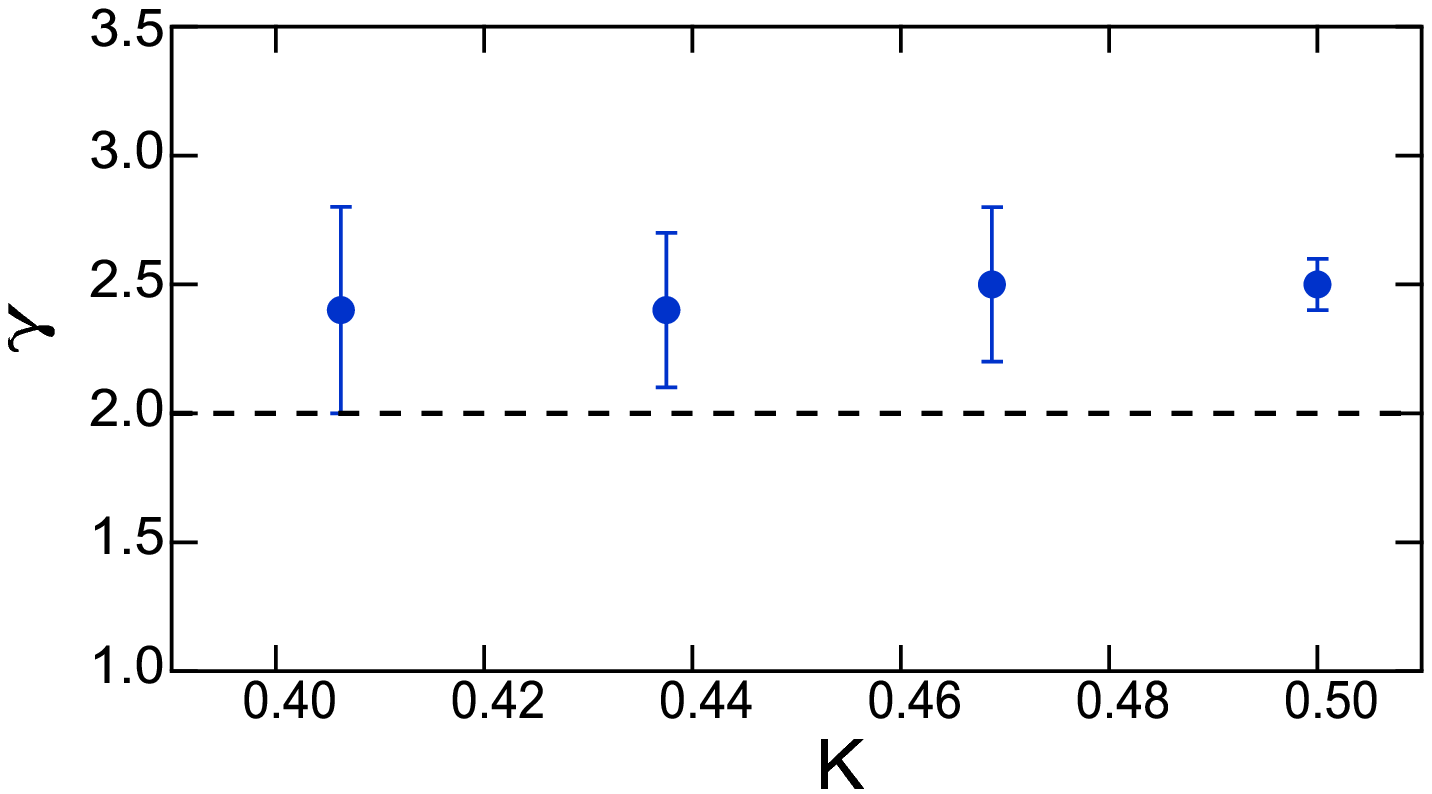} 
\vskip 0.1cm 
(c) \includegraphics[angle=0,width=0.93\columnwidth]{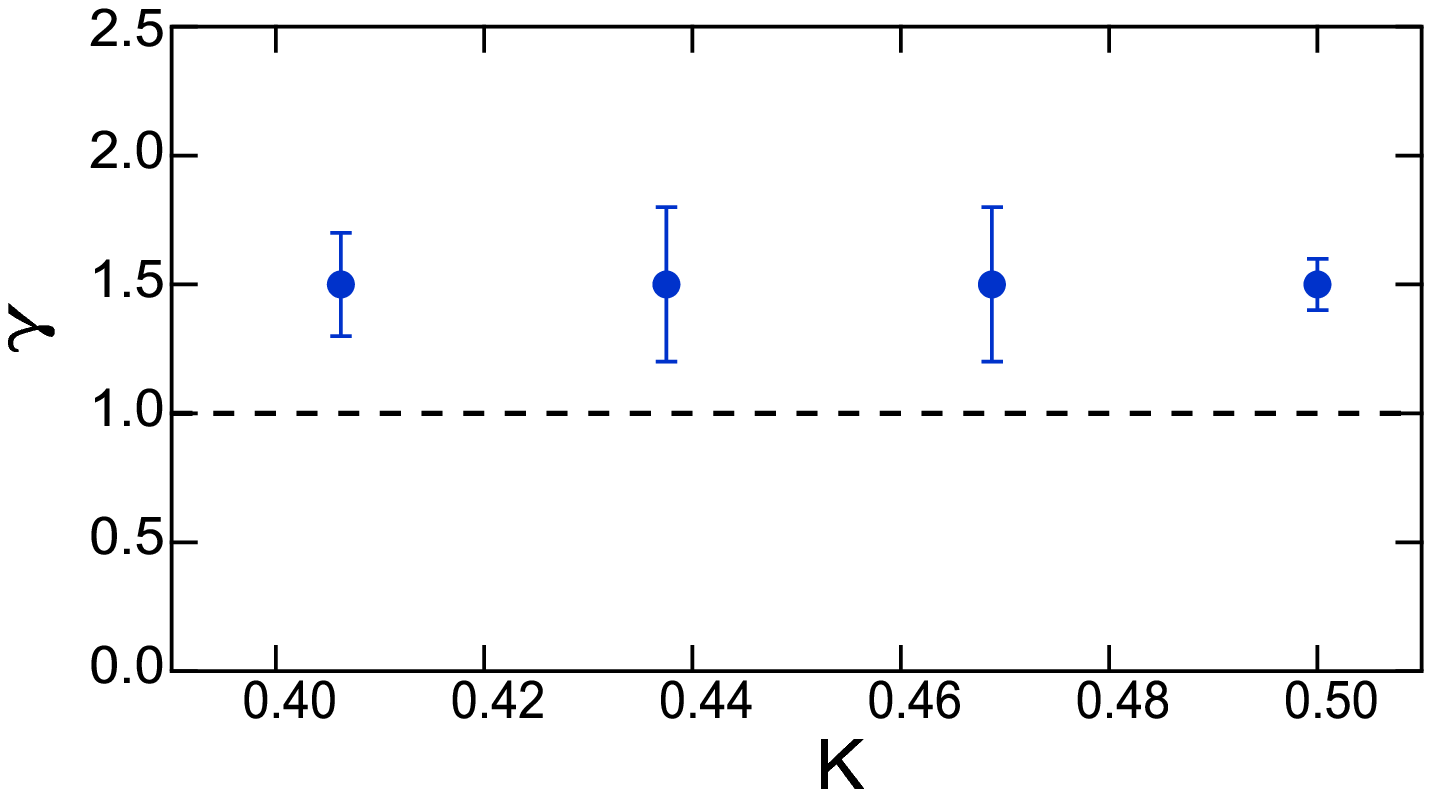}
\caption{(Color online) (a) Coulomb glass density of states $g(\epsilon)$ in
   three dimensions ($L = 8$, $N = 512$ sites) at temperature $T = 0.02$, 
   for various filling fractions $K = 0.5, 0.46875, 0.4375$, and $0.40625$ 
   (from bottom to top on right).
   (b) Corresponding (effective) gap exponents $\gamma(K)$ vs. filling fraction;
   (c) gap exponent $\gamma(K)$ in two dimensions ($L = 16$, $N = 256$, 
   $T = 0.02$); the dashed lines show the mean-field values $\gamma = d - 1$.} 
\label{fig2}
\end{figure}
At elevated temperature $T = 0.1$, $g(\mu_c)$ attains a non-zero value; when
the temperature scale reaches the width of the soft gap in the density of 
states, the Coulomb gap begins to fill owing to thermal excitations that 
wash out the sharp boundary between filled and empty energy levels.
As shown in Fig.~\ref{fig1}(b), the three-dimensional gap exponent $\gamma(T)$
becomes independent of $T$ once thermal excitations can be neglected. 
In contrast, in two dimensions, Fig.~\ref{fig1}(c), we observe a stronger 
temperature dependence of the effective gap exponent.
Extrapolating to $T \to 0$, our numerical values $\gamma \approx 2.5 \pm 0.2$
for $d = 3$ and $\gamma \approx 1.2 \pm 0.2$ deviate from the mean-field 
prediction $\gamma = d - 1$ (for $\sigma = 1$), and are in good agreement with 
Ref.~\cite{Mobius92}.
The displayed error bars merely represent the statistical errors which vary 
with the number of independent realizations used for each parameter set.

In Fig.~\ref{fig2}, we study the dependence of the soft Coulomb gap on the
filling fraction $K$.
Moving away from half-filling, $g(\epsilon)$ naturally becomes increasingly
asymmetric.
Yet near its minimum at $\mu_c$, the curves in Fig.~\ref{fig2}(a) collapse 
onto each other, resulting in gap exponents $\gamma(K)$ that are essentially 
independent of the total charge carrier density $K$, at least in the small
range $0.4 < K \leq 0.5$.

\begin{figure}
(a) \includegraphics[angle=0,width=0.93\columnwidth]{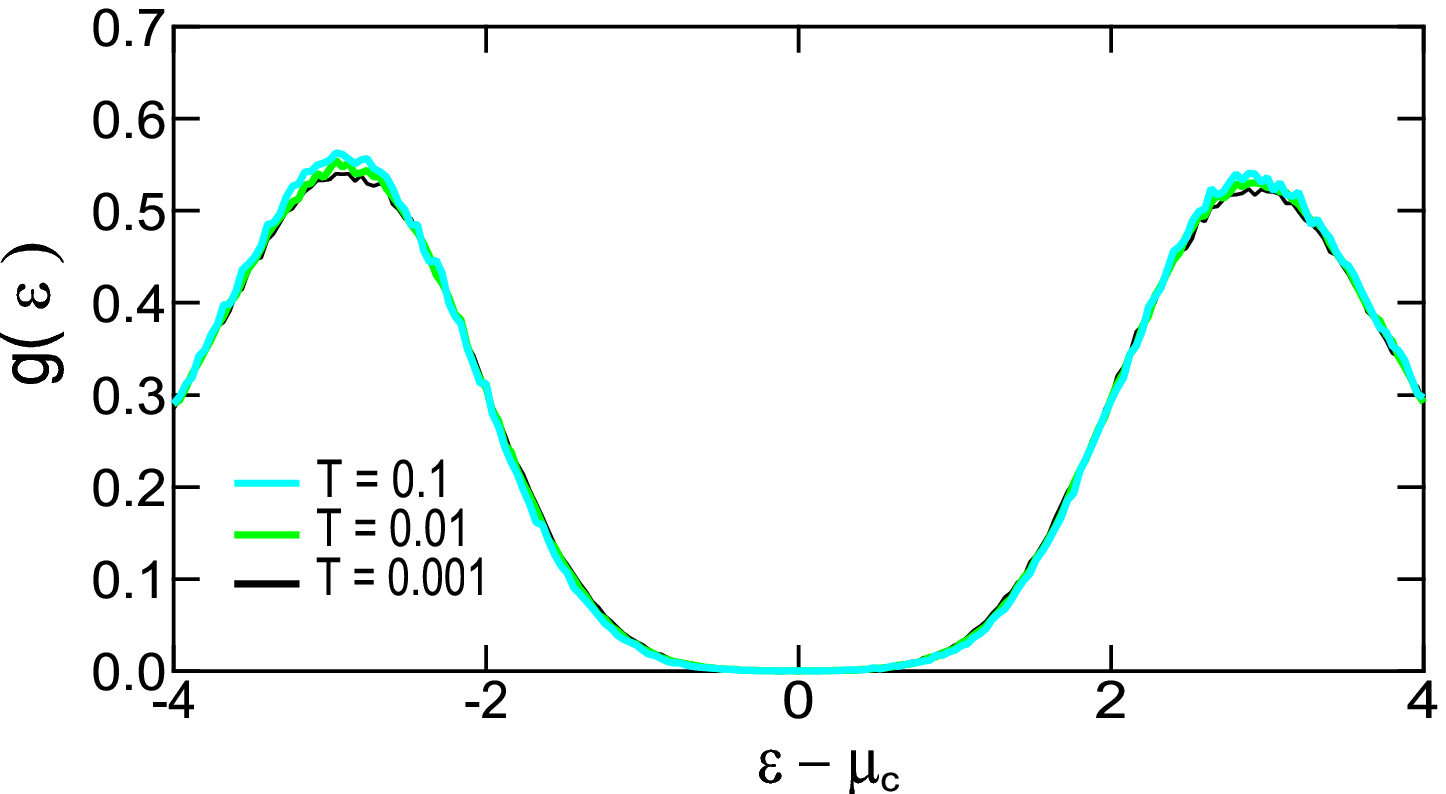} 
\vskip 0.1cm
(b) \includegraphics[angle=0,width=0.93\columnwidth]{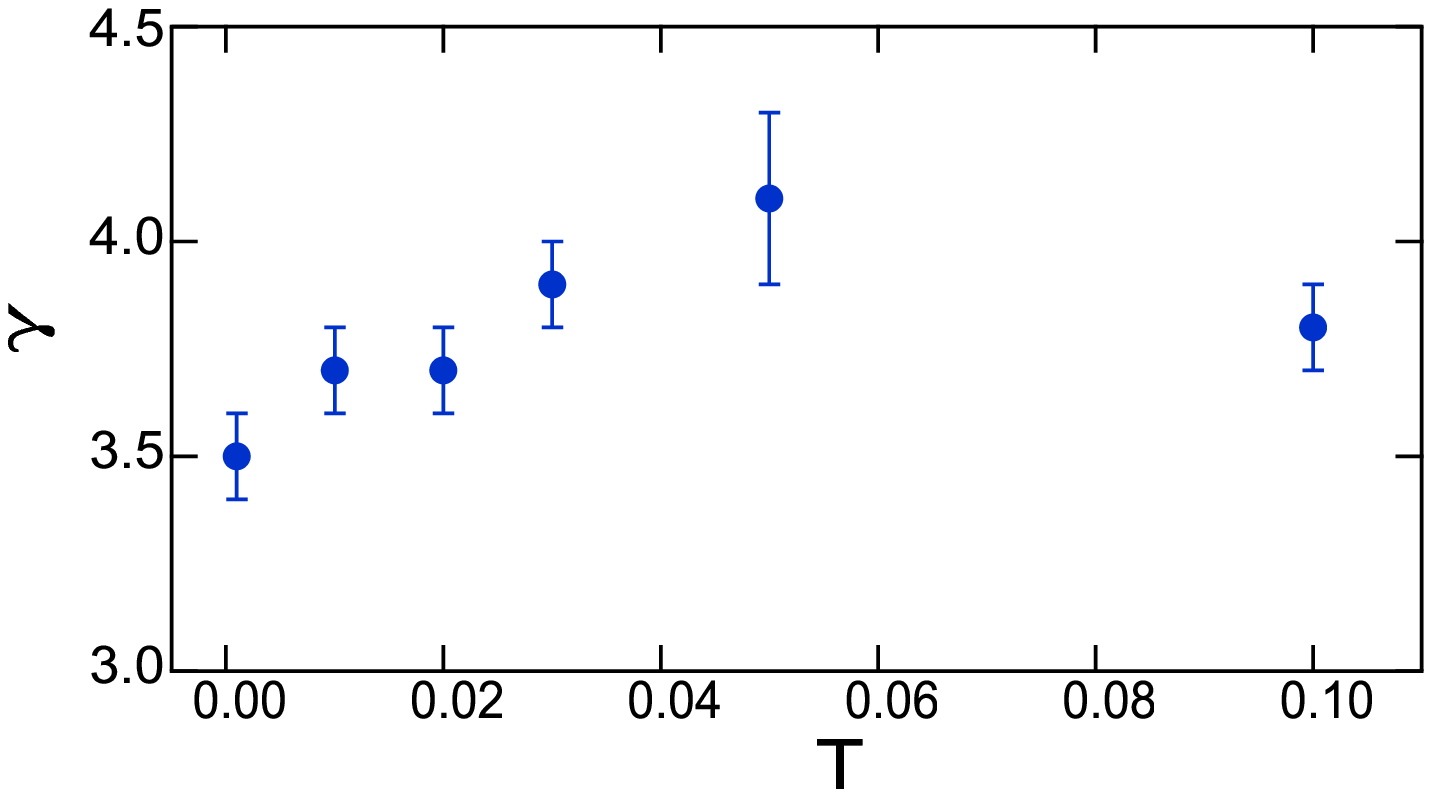}  
\caption{(Color online) (a) Density of states $g(\epsilon)$ for the 
   two-dimensional Bose glass ($L = 16$, $N = 256$ sites) for $K = 0.5$, at 
   temperatures $T = 0.001$ (black), $T = 0.01$ (green/gray), and $T = 0.1$ 
   (light blue/light gray).
   (b) Corresponding (effective) gap exponents $\gamma(T)$ vs. temperature.} 
\label{fig3}
\end{figure}
\begin{figure}
(a) \includegraphics[width=0.93\columnwidth]{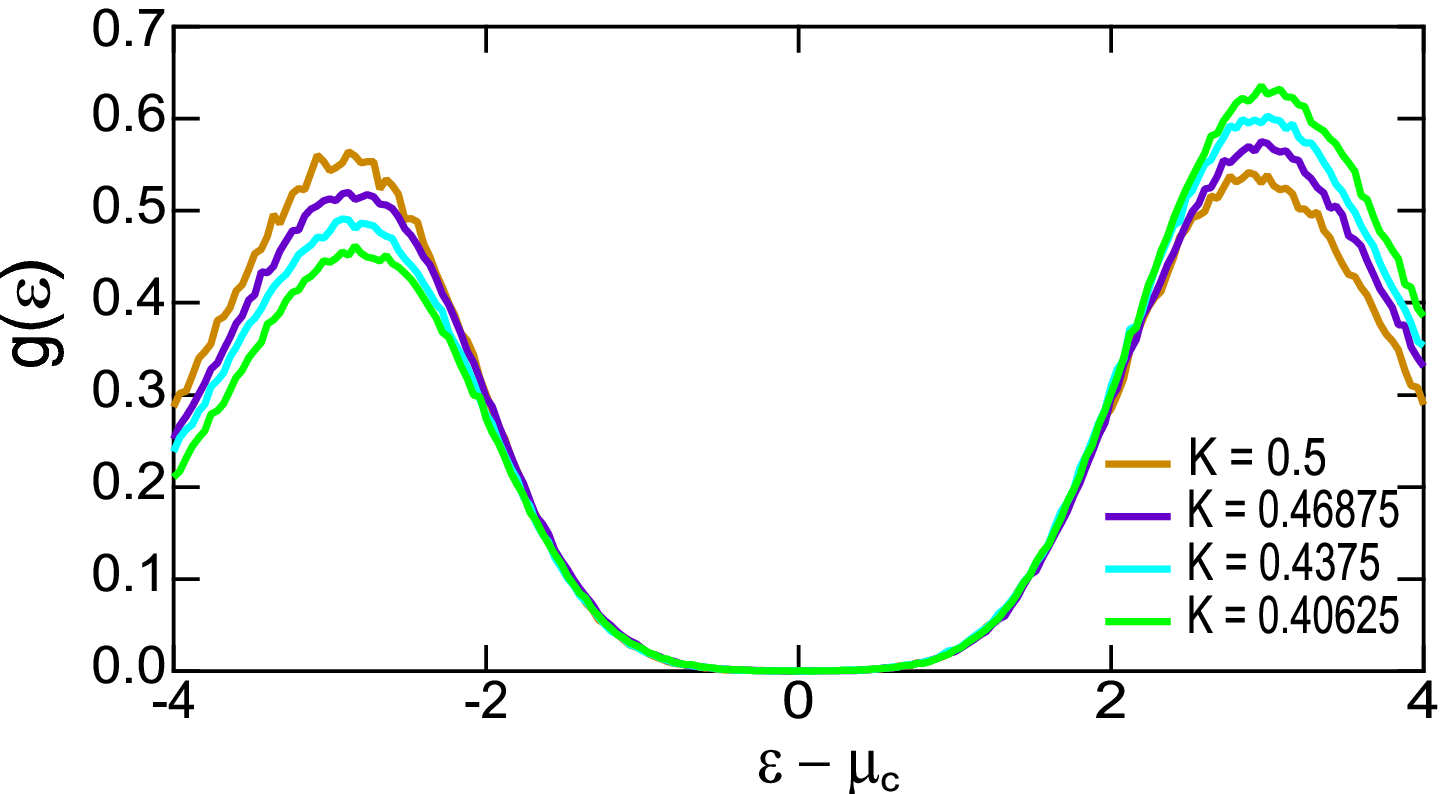} \vskip 0.1cm
(b) \includegraphics[width=0.93\columnwidth]{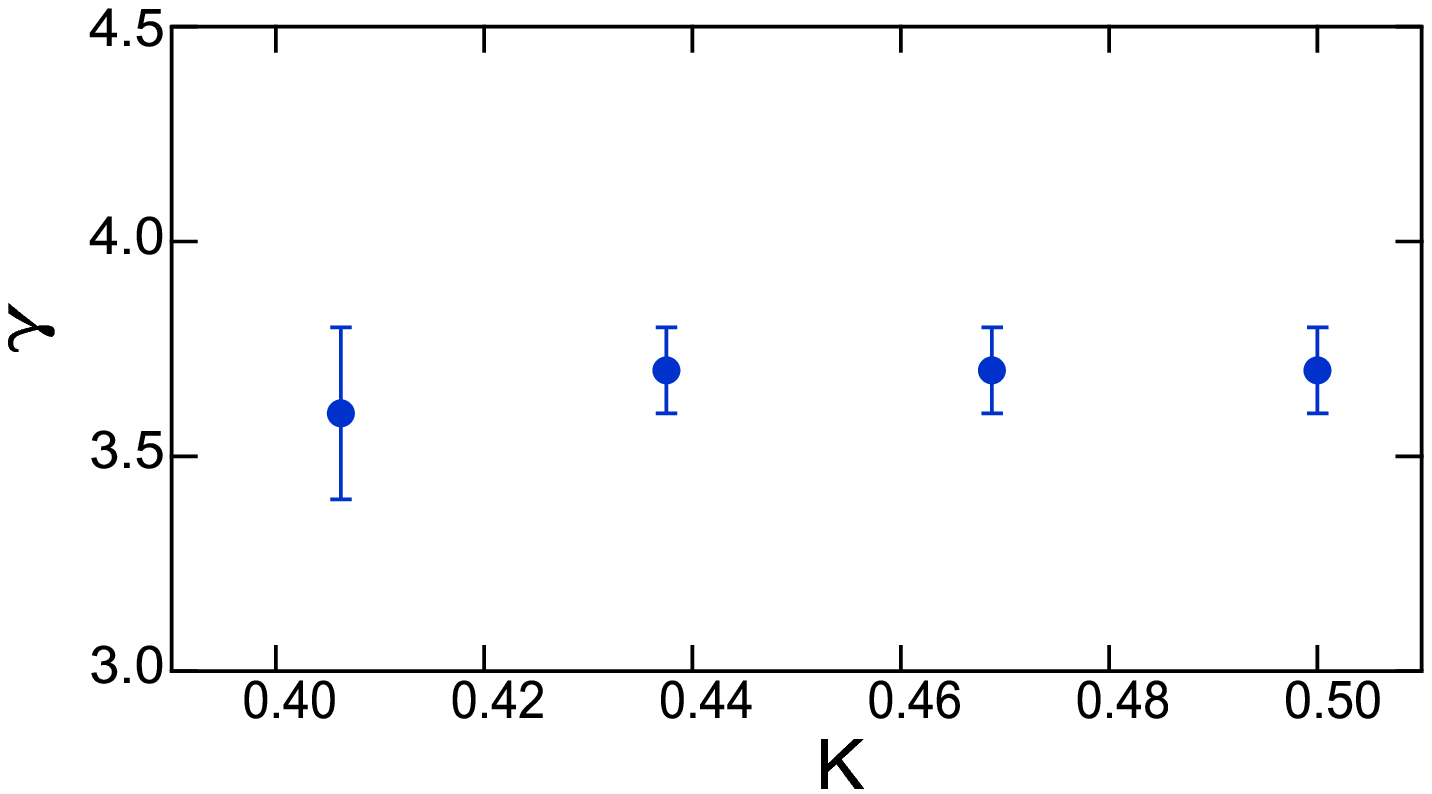}  
\caption{(Color online) Bose glass density of states $g(\epsilon)$ in two
   dimensions ($L = 16$, $N = 256$ sites) at temperature $T = 0.02$, for 
   filling fractions $K = 0.5, 0.46875, 0.4375$, and $0.40625$ (from bottom
   to top on right).
   (b) Corresponding (effective) gap exponents $\gamma(K)$ vs. filling 
   fraction.} 
\label{fig4}
\end{figure}

Next we explore the distribution of site energies in the two-dimensional 
Bose glass with long-range, essentially logarithmic repulsion 
($\lambda = 8 a_0$).
As is evident in Fig.~\ref{fig3}(a), the emerging soft correlation gap is
wider by a factor of $5$ as compared to the data for the Coulomb $1 / r$ 
interaction.
Therefore, even at $T = 0.1$ no thermal effects can be visibly discerned.
Yet measuring the effective gap exponent reveals an even steeper temperature 
dependence of $\gamma(T)$ than for the two-dimensional Coulomb glass, compare 
Fig.~\ref{fig3}(b) with Fig.~\ref{fig1}(c), extrapolating to 
$\gamma \approx 3.5 \pm 0.1$ as $T \to 0$ at half filling $K = 1/2$, a 
considerably larger value than reported in Ref.~\cite{Tauber95}; in that
study, however, no neutralizing charge background was employed, i.e., $K$ 
was set to zero in the Hamiltonian (\ref{bgla}).
As depicted in Fig.~\ref{fig4}, the dependence of the effective gap exponent
$\gamma(K)$ on the filling fraction $K$ is rather weak within the interval
$0.4 < K \leq 0.5$.

\section{Non-equilibrium Relaxation and Aging Scaling}
\label{aging}

We now proceed to our numerical results for non-equilibrium relaxation
features of the Coulomb and Bose glasses initially prepared in a random,
high-temperature state, as measured in the two-time density autocorrelation
function.
We first discuss the general relaxation scenario and the two distinct aging 
scaling fits we have implemented, before we provide the resulting scaling 
exponent values.

\subsection{Two-Time Density Autocorrelation Function}

\begin{figure*}
\centering (a) \qquad\qquad\qquad\qquad\qquad\qquad\qquad\qquad\qquad
\qquad\qquad\qquad\qquad (b) \\
\includegraphics[width=1.03\columnwidth]{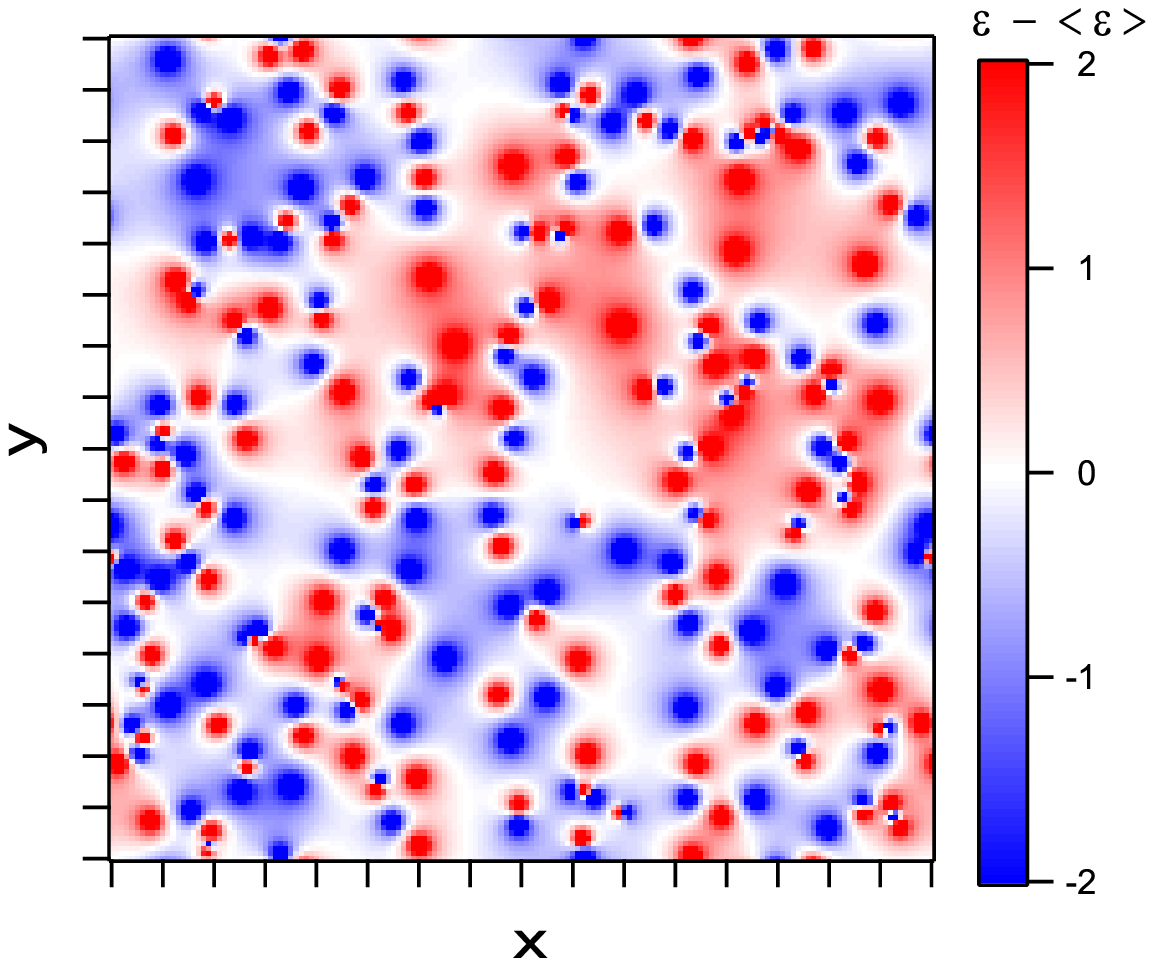} 
\includegraphics[width=1.03\columnwidth]{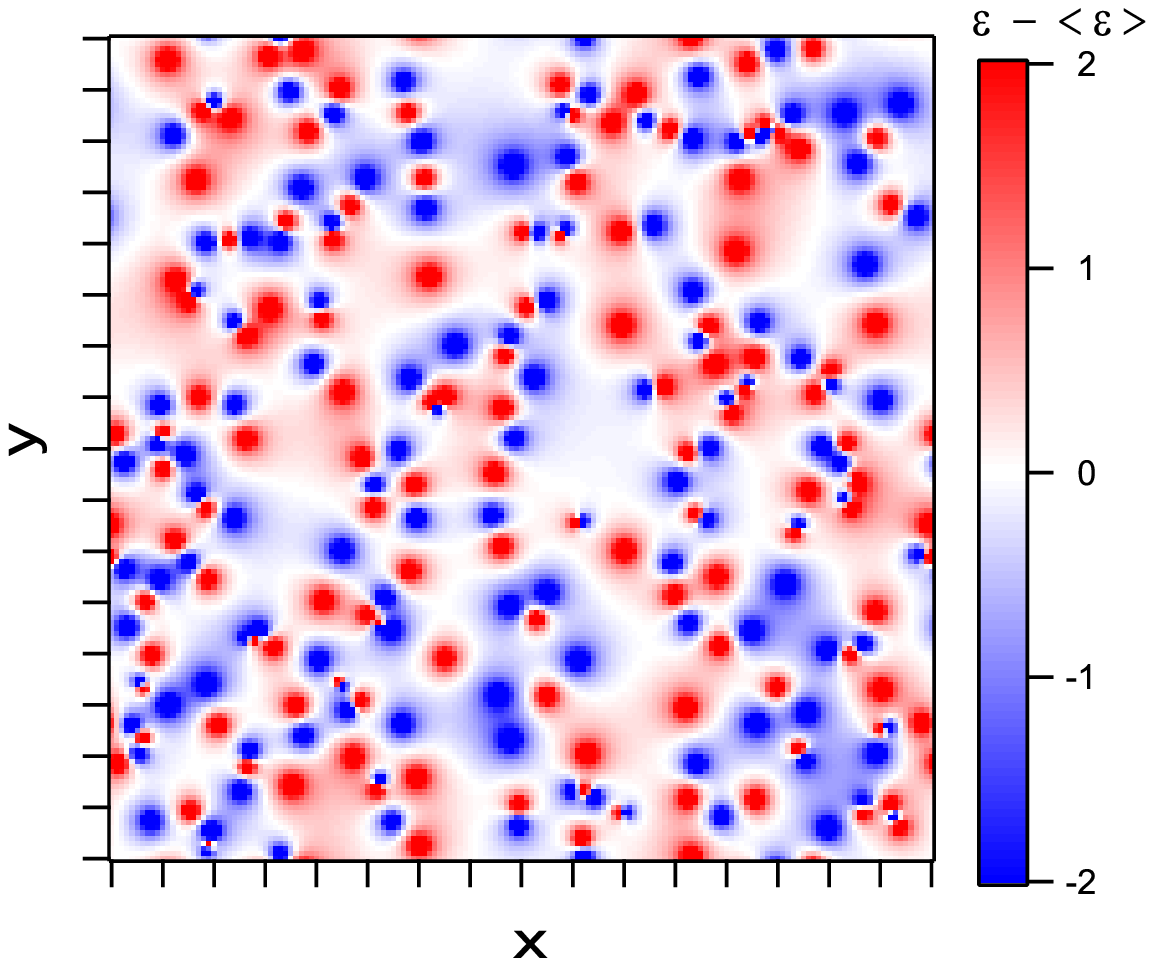}
 \caption{Energy landscape contour plots for the Coulomb glass with $1/r$ 
    potential in $d = 2$ dimensions ($L = 16$, $K = 1/2$, $T = 0.02$) after 
    (a) $10$ MCS and (b) $10^6$ MCS.}
\label{fig5}
\end{figure*}
In our initially entirely random distribution of charge carriers in the system,
inevitably many particles are placed in close vicinity.
They strongly repel each other and are fast displaced to energetically much
more favorable sites.
Correspondingly, the soft correlation-induced Coulomb gap in the density of 
states develops quite rapidly within $50 \ldots 100$ MCS.
Subsequently subtle spatial rearrangements take place that further reduce 
the total energy, as becomes clearly visible in the temporal evolution of the
energy landscape contour plots shown in Fig.~\ref{fig5}.
These processes proceed on considerably longer time scales; yet in this 
intermediate regime the system retains memory of its initial configuration, 
and time translation invariance is broken, in contrast to the asymptotic 
stationary, equilibrated state \cite{Henkel07, Henkel09}.

In order to monitor the slow structural relaxation kinetics in the Coulomb
and Bose glass, we compute the (normalized) two-time carrier density 
autocorrelation function \cite{Grempel04, Kolton05, Shimer10}
\begin{equation}
\label{auto}
   C(t,s) = \frac{\langle n_i(t) n_i(s) \rangle - K^2}{K (1 - K)}
   = \frac{\sum_i n_i(t) n_i(s) - K^2 N}{K (1 - K) N} \, ,
\end{equation}
where $s$ indicates the elapsed time after the high-tem\-perature quench, when
the Monte Carlo simulation runs are initiated, while $t > s$ refers to a later
``measurement time'' when the temporal correlations are obtained relative to 
the ``waiting time'' $s$.
Since $n_i^2 = n_i$ and $\sum_i n_i = K N$, at equal times $C(s,s) = 1$.

\begin{figure}[b]
(a) \includegraphics[width=0.93\columnwidth]{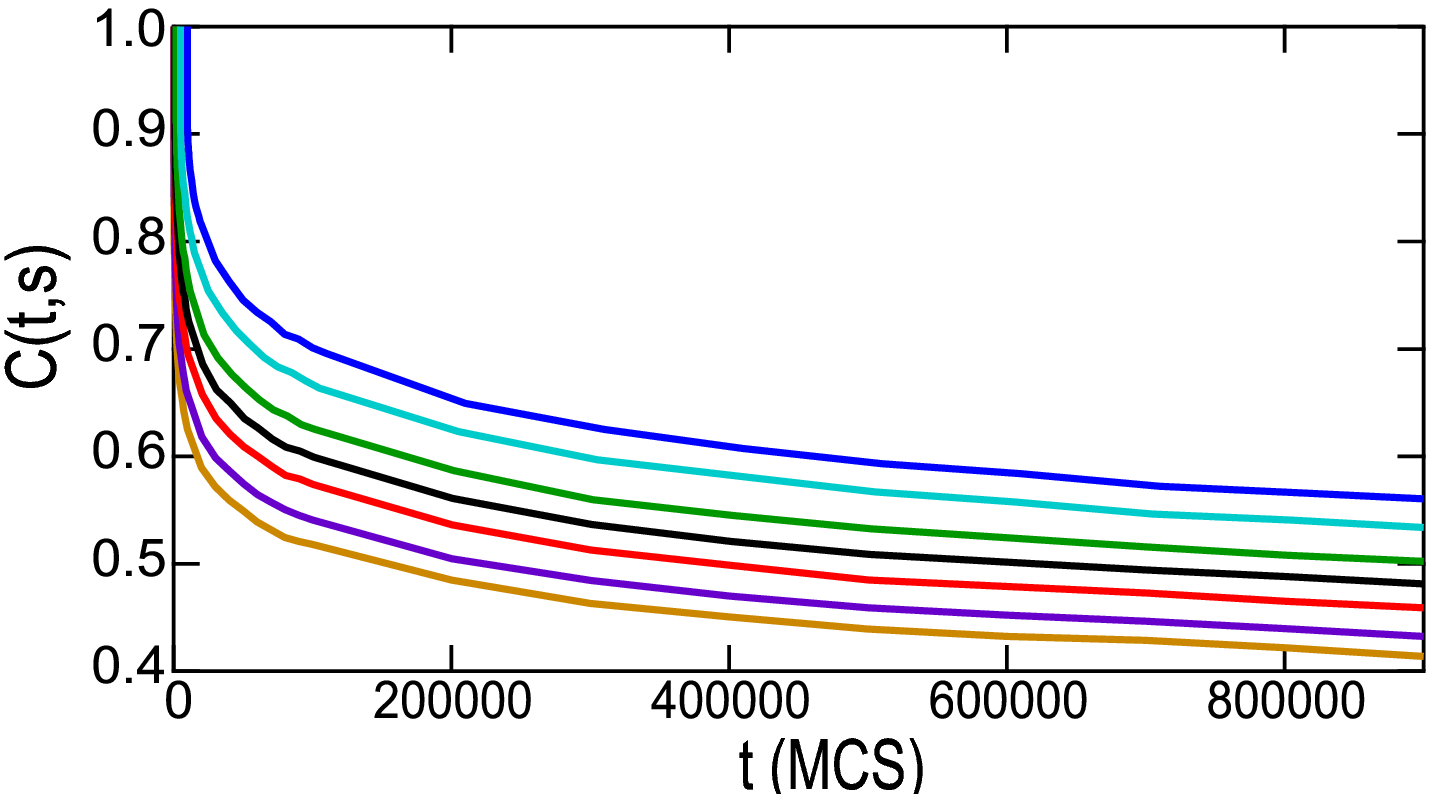} \vskip 0.2cm
(b) \includegraphics[width=0.93\columnwidth]{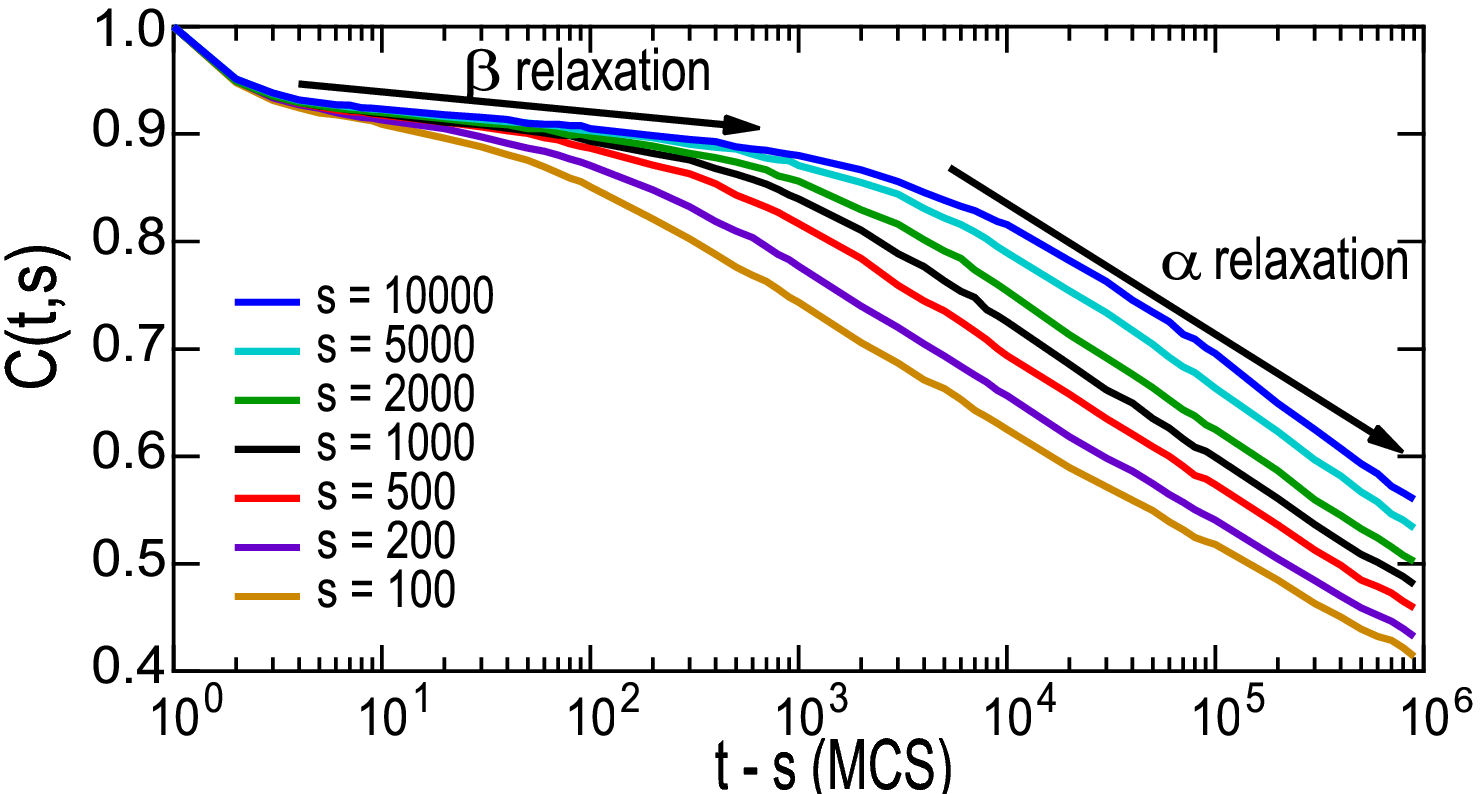}  
\caption{(Color online) Non-equilibrium relaxation and aging for the carrier
   density autocorrelation function (\ref{auto}) in the two-dimensional
   Coulomb glass (for $L = 16$, $K = 1/2$, $T = 0.02$).
   (a) $C(t,s)$ vs. $t$ for various waiting times
   $s = 100, 200, 500, 1000, 2000, 5000, 10000$ (from bottom to top); 
   (b) same data plotted vs. $t - s$ on a logarithmic scale.} 
\label{fig6}
\end{figure}
Representative data from our simulation runs are displayed in Fig.~\ref{fig6}.
The linear plot of $C(t,s)$ vs. the measurement time $t$ in Fig.~\ref{fig6}(a)
shows that even after $10^6$ MCS no stationary, equilibrium state has been 
reached yet.
Graphing the same autocorrelation data against the time difference $t - s$, in
Fig.~\ref{fig6}(b) on a logarithmic scale, establishes that time translation
invariance is indeed manifestly broken.
In accord with the data of Ref.~\cite{Grempel04}, we observe that following a 
fast initial decay towards an almost flat quasi-``plateau'' region, the graphs 
for different waiting times $s$ become distinct. 
Indeed, the longer ``aged'' runs for larger waiting times remain in an
intermediate state for more extended time periods, before the density 
autocorrelation ultimately resumes its slower relaxation towards zero.
In analogy with the phenomenology in structural glasses (see, e.g., 
Ref.~\cite{Gotze92}), we term these two distinct relaxation regimes visible in
our data ``$\beta$'' and ``$\alpha$ relaxation'', respectively.
In the following, we address the power law scaling for the slow density
relaxation processes in the $\alpha$ relaxation regime.

\subsection{Dynamical Aging Scaling}

\begin{figure}
(a) \includegraphics[width=0.93\columnwidth]{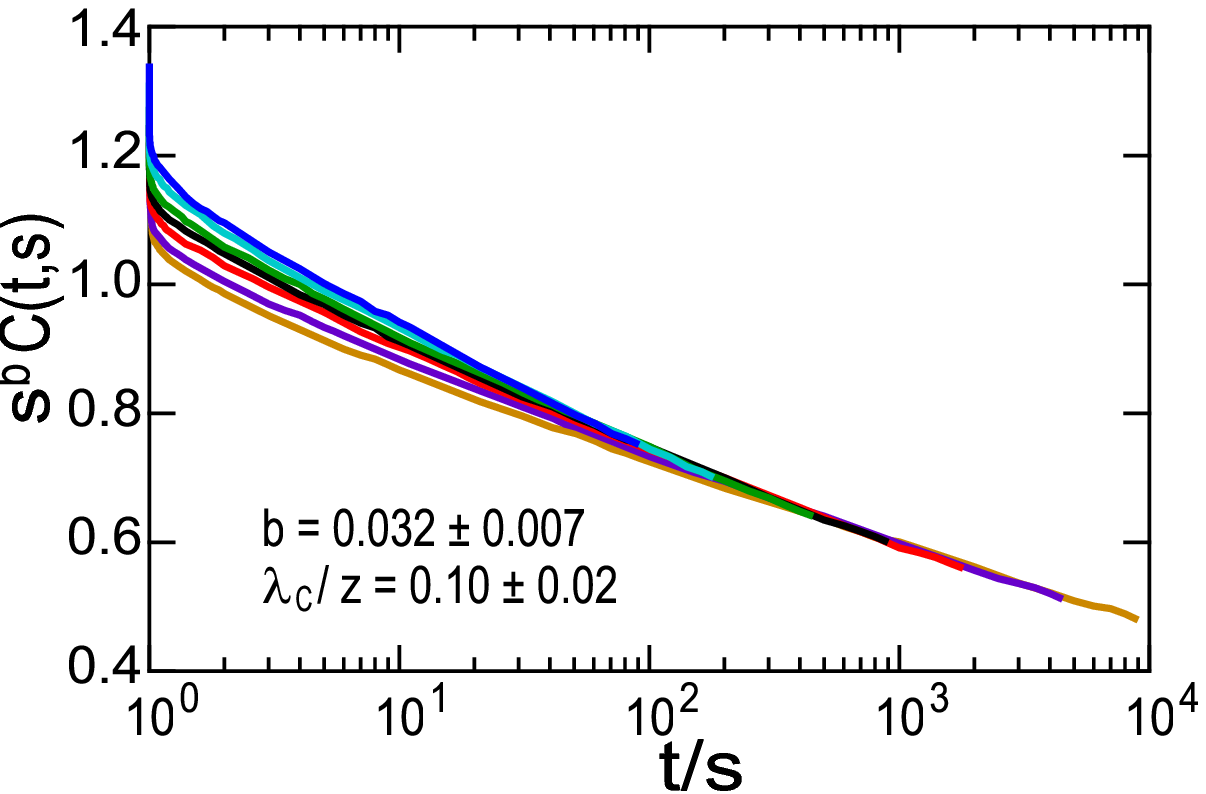} \vskip 0.1cm
(b) \includegraphics[width=0.93\columnwidth]{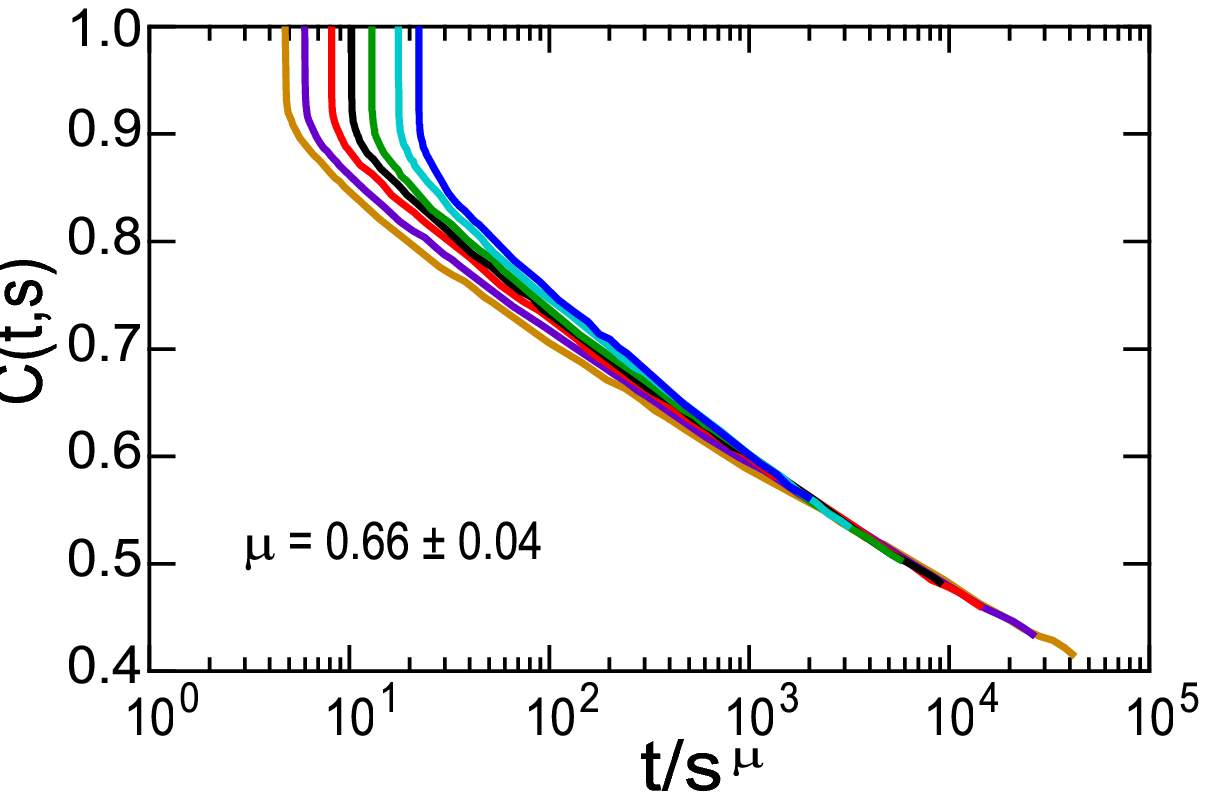}  
\caption{(Color online) Aging scaling collapse for the density autocorrelation
   function (\ref{auto}) in the two-dimensional Coulomb glass (for $L = 16$, 
   $K = 1/2$, and $T = 0.02$), obtained for the set of waiting times 
   $s = 100, 200, 500, 1000, 2000, 5000, 10000$ (from bottom to top):
   (a) full aging scaling according to Eq.~(\ref{gags}) with $\mu = 1$;
   (b) subaging scaling, Eq.~(\ref{gags}) with $b = 0$.} 
\label{fig7}
\end{figure}
We consider the aging scaling limit, where both $s, t \gg \tau_0$ (or any other
microscopic time scale), and in addition $t \gg s$. i.e., $t - s \gg \tau_0$.
In the $\alpha$ relaxation regime, time translation invariance does not hold, 
whence the carrier density two-time autocorrelation function (\ref{auto}) does 
not just depend on the time difference $t - s$, but on both $t$ and $s$ 
separately.
Following the notations in Ref.~\cite{Henkel09}, we posit the following general
aging scaling form
\begin{equation}
\label{gags}
   C(t,s) = s^{- b} f_C(t / s^\mu) \, ,
\end{equation}
with scaling exponents $b \geq 0$ and $\mu \leq 1$.
In many simple, analytically tractable situations, characterized by a single
algebraically growing length scale $L(t) \sim t^{1/z}$ with dynamic scaling 
exponent $z \geq 1$, one in fact obtains aging scaling laws of the form 
(\ref{gags}) with $\mu = 1$, often referred to as ``full aging''.
In the limit $t / s \to \infty$, in this situation one furthermore expects the 
scaling function to follow the algebraic decay 
\begin{equation}
\label{scfd}
  f_C(x) \sim [L(t) / L(s)]^{- \lambda_C} \sim (t / s)^{- \lambda_C / z}
\end{equation}
with the autocorrelation exponent $\lambda_C \geq 0$.
Prominent examples that display this full aging scaling scenario are the purely 
relaxational dynamics in the kinetic Ising model in one dimension 
\cite{Godreche00, Lippiello00}, time-dependent Ginzburg-Landau models 
quenched to the critical point \cite{Janssen89, Calabrese05}, and coarsening of 
the spherical model A in the low-temperature phase, both with short-range 
\cite{Coniglio90, Biroli05} and long-range \cite{Cannas01, Baumann07, Dutta08} 
interactions.

In Fig.~\ref{fig7}(a), we demonstrate scaling collapse of the two-dimensional
Coulomb glass density autocorrelation data from Fig.~\ref{fig6} ($L = 16$) 
utilizing the full-aging scaling form (\ref{gags}) with $\mu = 1$.
Focusing on the data for $t - s$ in the range $3 \cdot 10^4 \ldots 10^6$ MCS,
and following the interpolation method described in 
Ref.~\cite{Bhattacharjee01}, we obtain optimal collapse onto a single master 
curve for $b = 0.032 \pm 0.007$.
From the asymptotic long-time decay we furthermore infer 
$\lambda_C / z \approx 0.10 \pm 0.02$ for $T = 0.02$ at half filling 
$K = 1/2$ \cite{Shimer10}.

\begin{figure}
(a) \includegraphics[width=0.93\columnwidth]{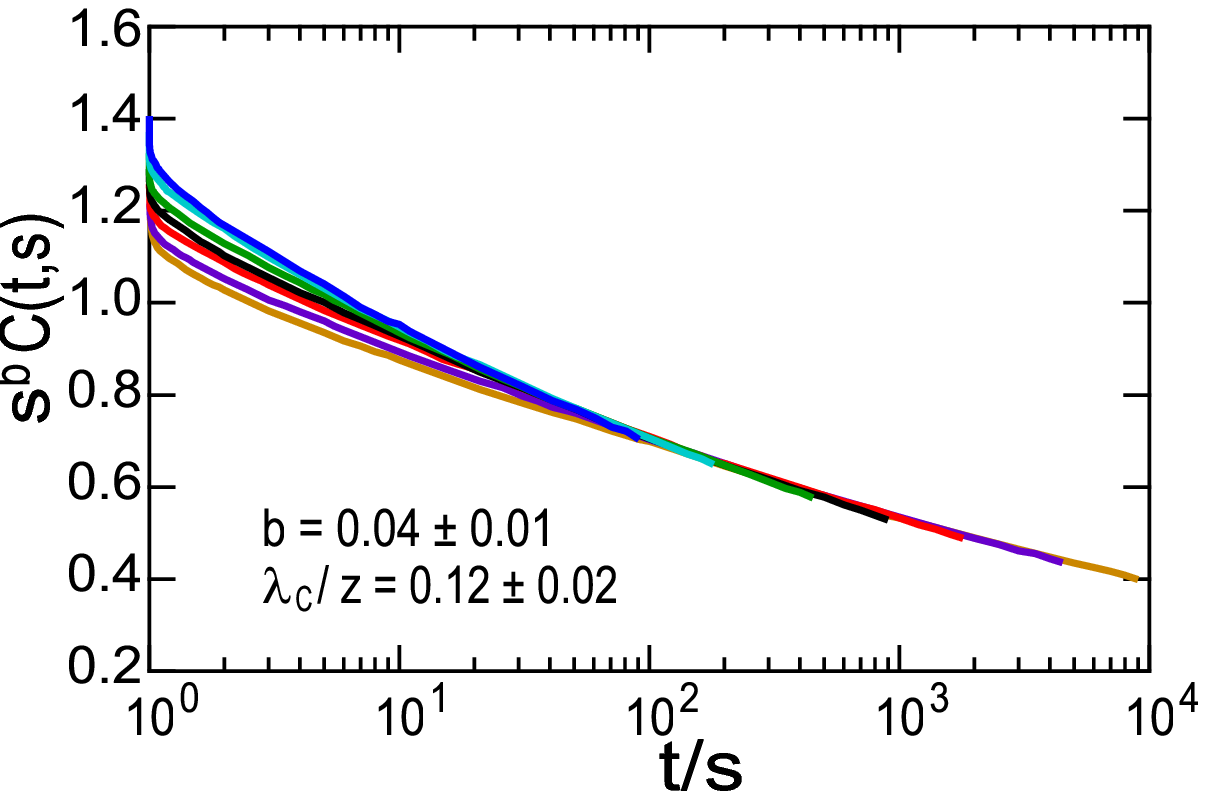} \vskip 0.1cm
(b) \includegraphics[width=0.93\columnwidth]{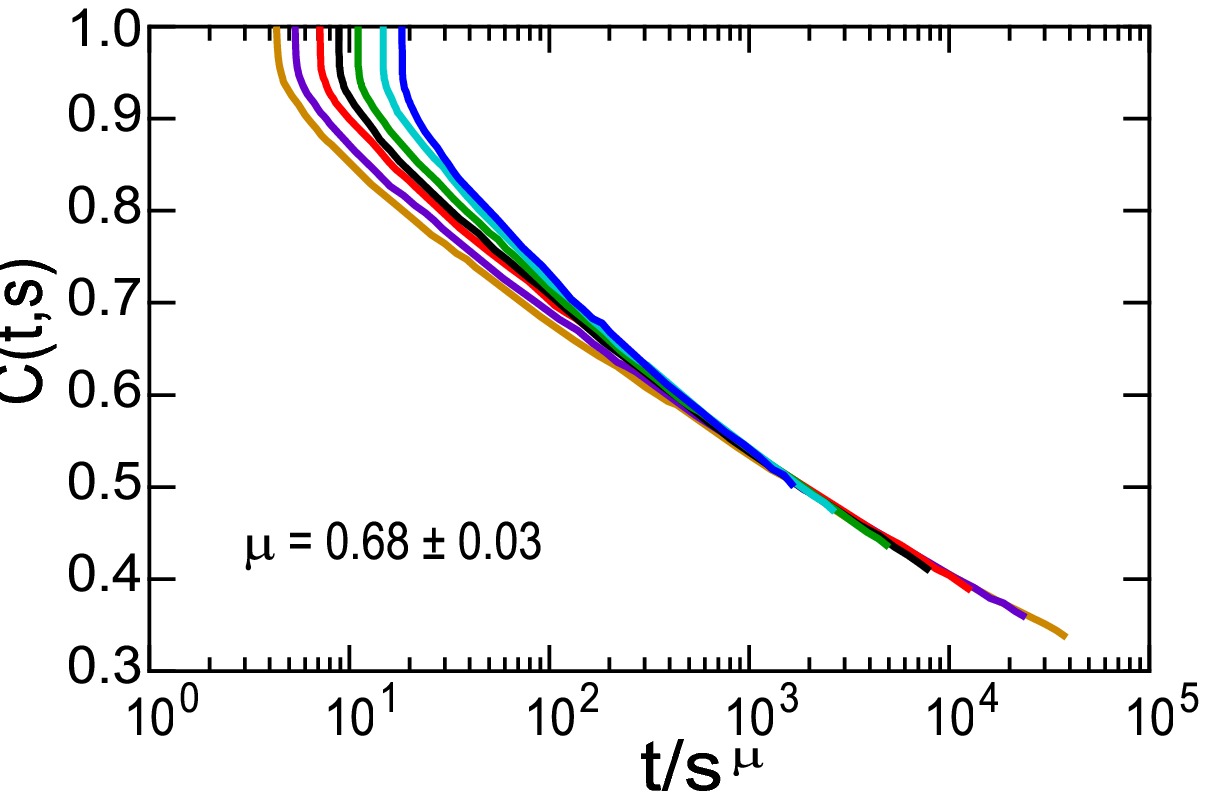}  
\caption{(Color online) Aging scaling collapse for the density autocorrelations
   as in Fig.~\ref{fig7}, but now for the three-dimensional Coulomb glass
   ($L = 8$, $K = 1/2$, and $T = 0.02$; waiting times $s$ as in 
   Figs.~\ref{fig6} and \ref{fig7}).}
\label{fig7a}
\end{figure}
\begin{figure}
(a) \includegraphics[width=0.93\columnwidth]{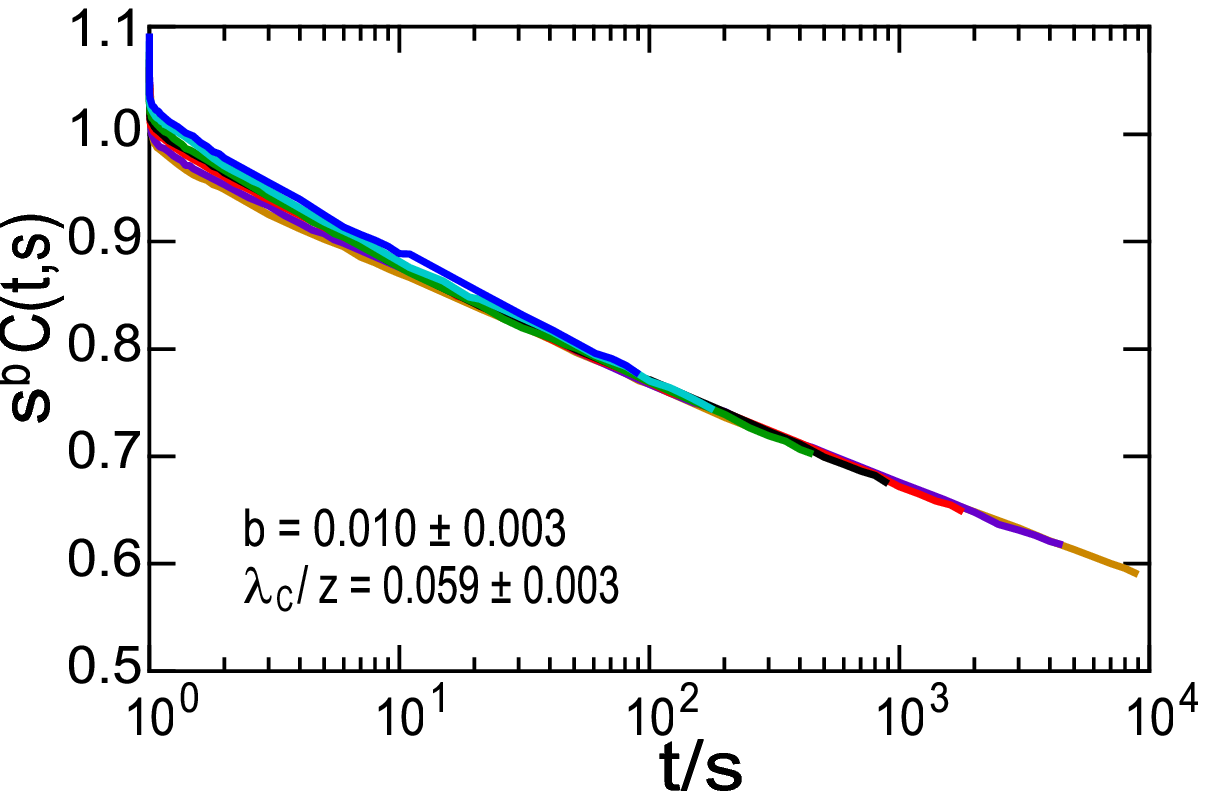} \vskip 0.1cm
(b) \includegraphics[width=0.93\columnwidth]{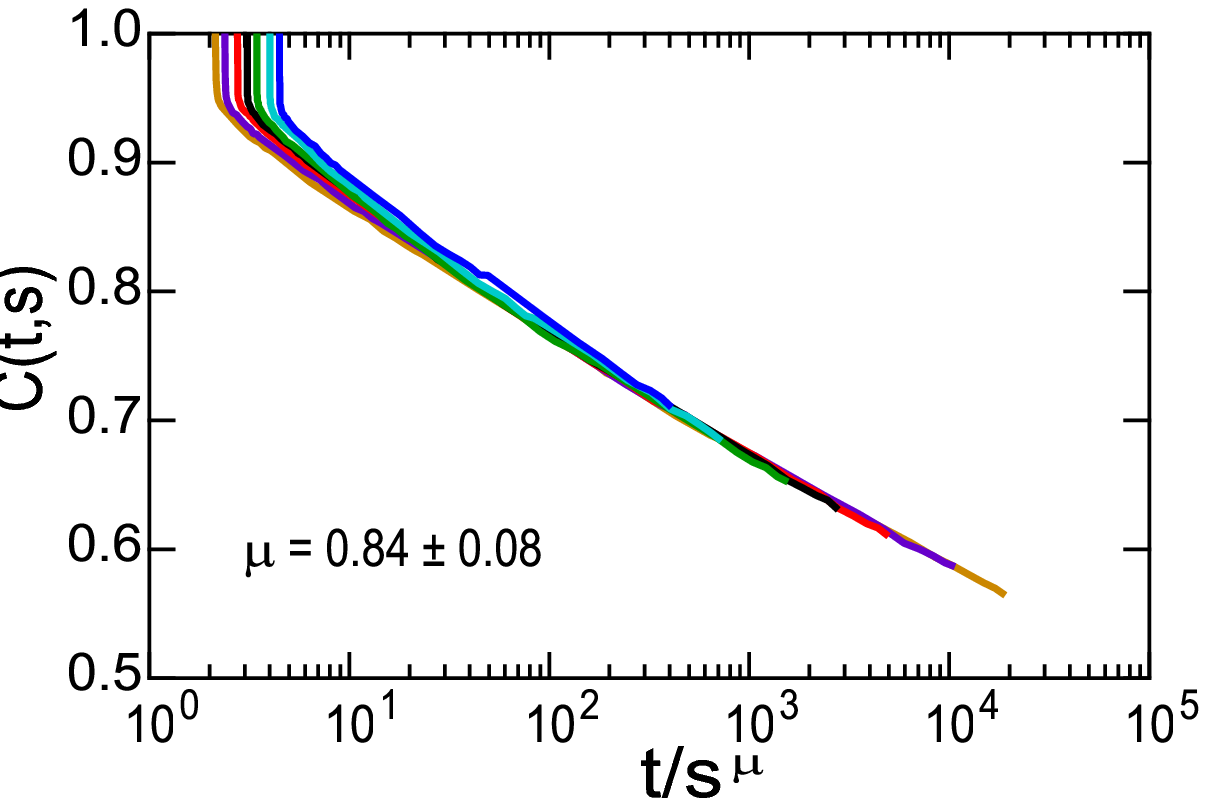}  
\caption{(Color online) Aging scaling collapse for the density autocorrelations
   as in Fig.~\ref{fig7}, for the two-dimensional Bose glass with $L = 24$, 
   $K = 1/2$, and $T = 0.02$; waiting times $s$ as in Figs.~\ref{fig6} and 
   \ref{fig7}.}
\label{fig7b}
\end{figure}

Alternatively, one may impose $b = 0$ in Eq.~(\ref{gags}), and instead work 
with a non-trivial scaling exponent $\mu < 1$; this ``subaging scaling'' is 
frequently employed in the spin glass literature \cite{Henkel09}.
As purely phenomenological fits, both scaling ans\"atze are in essence
equivalent.
The subaging scaling collapse of our data for the two-dimensional Coulomb glass
is depicted in Fig.~\ref{fig7}(b), best fit with the value 
$\mu = 0.66 \pm 0.04$ \cite{Shimer10}.
Similar scaling properties are obtained for the three-dimensional Coulomb glass
as well as for the two-dimensional Bose Glass. 
Figs.~\ref{fig7a} and \ref{fig7b} show a characteristic example for each of 
these two cases.

\subsection{Coulomb / Bose Glass Aging Scaling Exponents}

We collected data for the two-time density autocorrelation function for the 
Coulomb glass model in two (with $L = 16$, $N = 256$) and three dimensions 
($L = 8$, $N = 512$), as well as for the two-dimensional Bose glass with 
essentially logarithmic repulsion ($L = 24$, $N = 576$) at various 
temperatures and filling fractions.
We then determined the associated scaling exponents in the long-time aging 
regime following the procedures outlined in the previous subsection.
(The resulting scaling plots can be found in Ref.~\cite{Shimer11}.)

\begin{figure}
(a) \includegraphics[width=0.93\columnwidth]{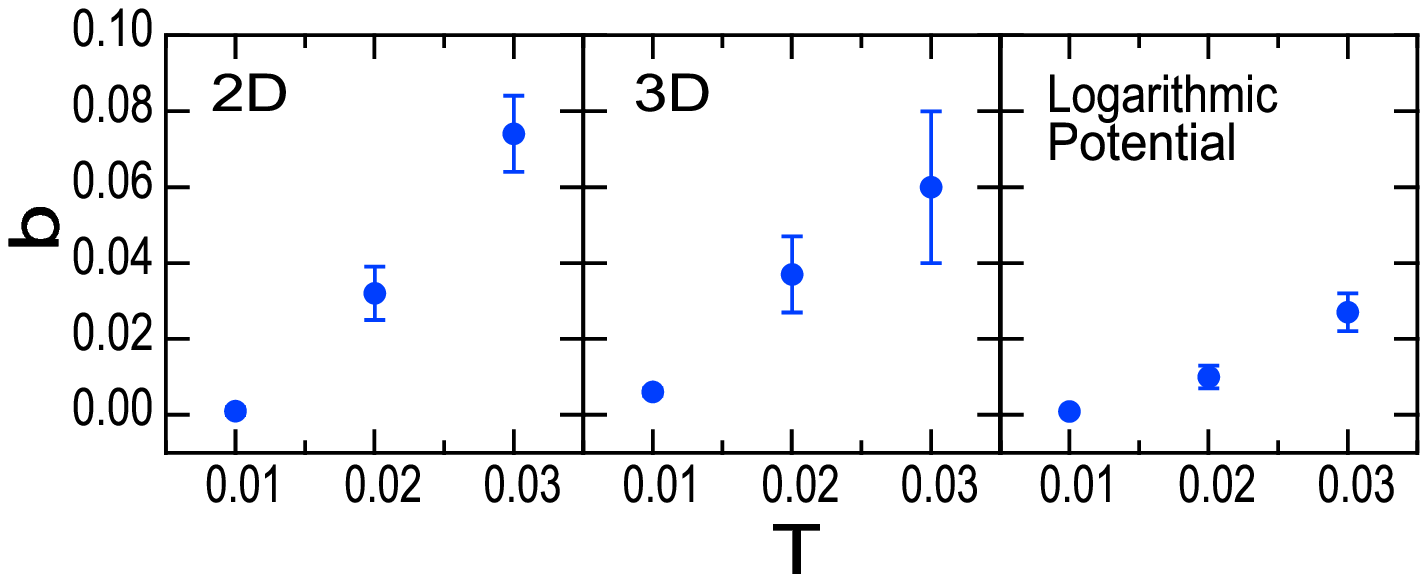} \vskip 0.1cm
(b) \includegraphics[width=0.93\columnwidth]{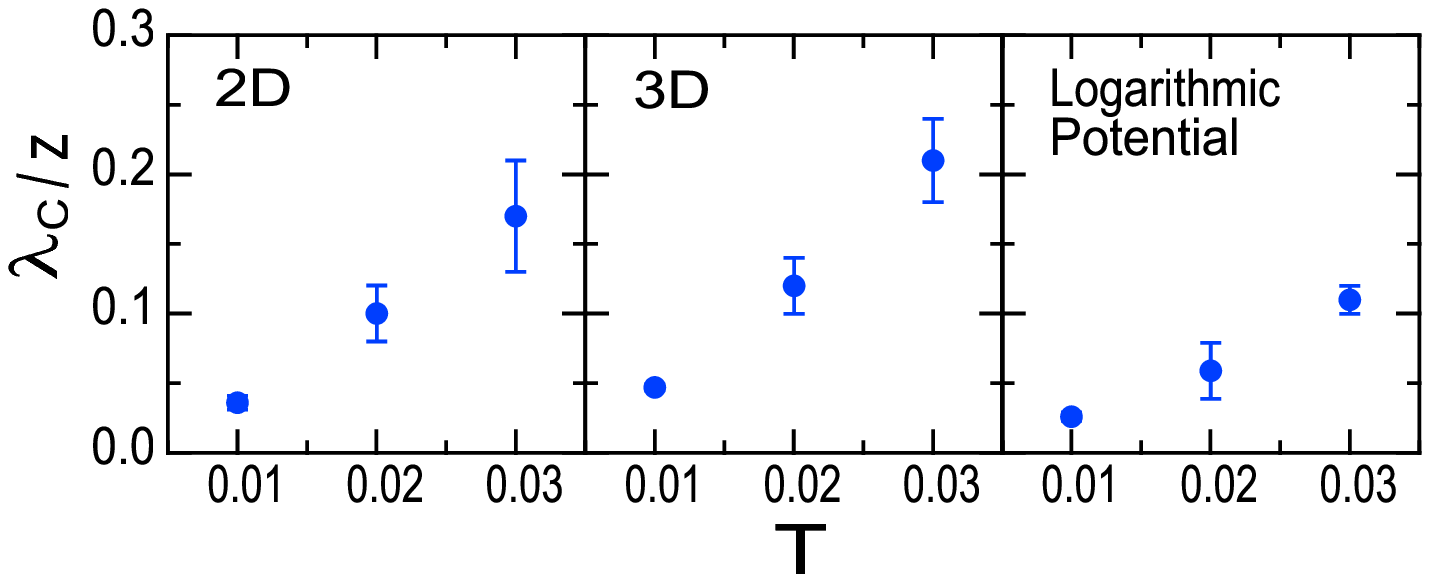} \vskip 0.1cm
(c) \includegraphics[width=0.93\columnwidth]{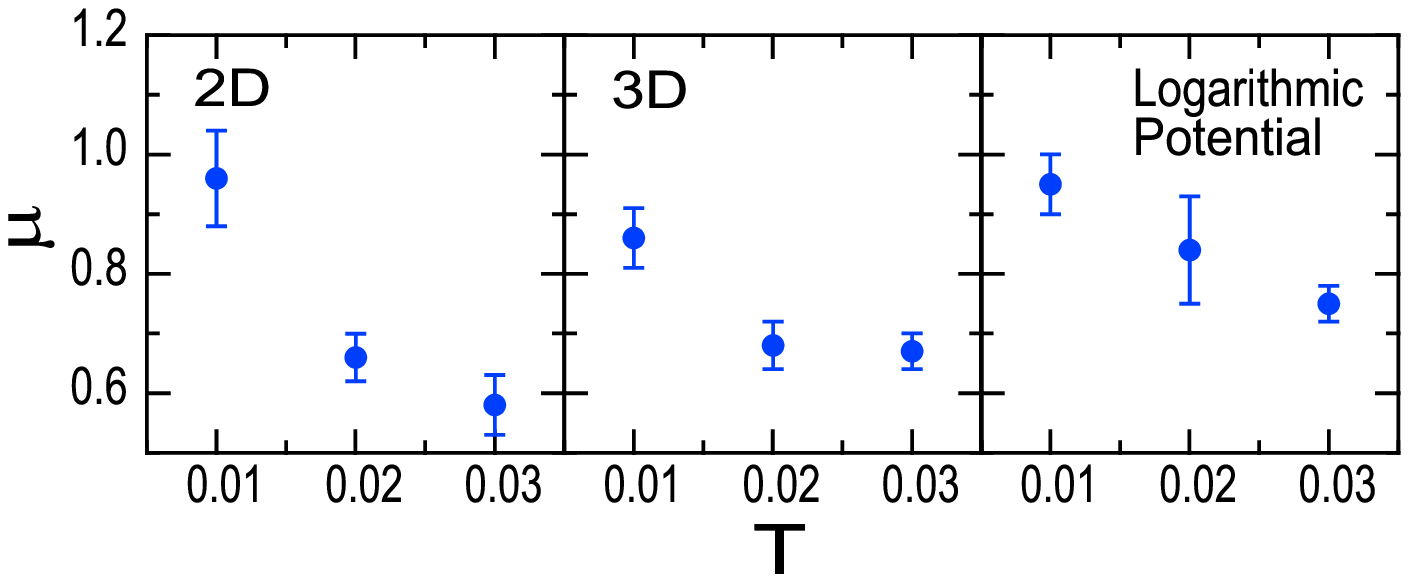}
\caption{Aging scaling exponents at $K = 1/2$ as functions of $T$ for the
   Coulomb glass ($1/r$ interaction) in two (left panels) and three dimensions
   (center), and for the Bose glass (with logarithmic repulsion, right):
   (a) full aging scaling exponent $b$, Eq.~(\ref{gags}) with $\mu = 1$;
   (b) autocorrelation decay exponent $\lambda_C / z$, Eq.~(\ref{scfd});
   (c) subaging exponent $\mu$, Eq.~(\ref{gags}) with $b = 0$.} 
\label{fig8}
\end{figure}

The thus obtained full-aging scaling exponents $b$ and $\lambda_C / z$,
see Eqs.~(\ref{gags}) with $\mu = 1$ and (\ref{scfd}), are compiled in
Fig.~\ref{fig8}(a) and (b) at half filling $K = 1/2$ as functions of the 
temperature $T$ (see also Refs.~\cite{Grempel04, Kolton05}).
As one would expect, the non-equilibrium relaxation from the randomized
initial state slows down drastically upon lowering the temperature, here
clearly reflected in successively smaller values for $b$ and $\lambda_C / z$
as $T$ is reduced from $0.03$ to $0.01$.
Indeed, for even lower temperatures $T < 0.01$, our systems basically freeze
in and we could not obtain statistically meaningful data for the ensuing
extremely rare relaxation events.
At $T = 0.01$, we find $b = 0.001 \pm 0.001$ for the two-dimensional 
Coulomb glass, see the left panel in Fig.~\ref{fig8}(a), borderline consistent
with the recently developed mean-field theory for aging relaxation in 
disordered electron glasses that predicts logarithmic scaling 
\cite{Amir08, Amir09, Oreg09, Amir11}.
However, in three dimensions we measure $b = 0.006 \pm 0.001$ at our lowest
accessible temperature $T = 0.01$ (center panel), while for the 
two-dimensional Bose glass $b = 0.0009 \pm 0.0003$ (right panel).
The associated autocorrelation to dynamic exponent ratios at $T = 0.01$ are
$\lambda_C / z = 0.036 \pm 0.005$ (Coulomb glass, $d = 2$), 
$\lambda_C / z = 0.047 \pm 0.002$ (Coulomb glass, $d = 3$), and 
$\lambda_C / z = 0.026 \pm 0.003$ (Bose glass, $d = 2$), 
see Fig.~\ref{fig8}(b).
Note that relaxation processes in the Bose glass generically happen much
slower as compared to the Coulomb glass (in $d = 2$ and $d = 3$ dimensions),
as a consequence of the much shallower soft gap in the density of states, see
Figs.~\ref{fig2}(a) and \ref{fig4}(a).  

Our corresponding results from the alternative subaging scaling analysis, 
Eq.~\ref{gags} with $b = 0$, are plotted in Fig.~\ref{fig8}(c).
Note that the drastic slowing-down of the relaxation processes with reduced
temperature now becomes apparent as a marked increase of the subaging scaling
exponent $\mu$, which almost approaches $1$ for the two-dimensional Coulomb
glass at $T = 0.01$.
At this lowest temperature and half filling $K = 1/2$, our data yield 
$\mu = 0.96 \pm 0.008$ and $\mu = 0.86 \pm 0.05$ for the Coulomb glass in two
and three dimensions, respectively, and $\mu = 0.9 \pm 0.0$ for the
two-dimensional Bose glass.

\begin{figure}
(a) \includegraphics[width=0.93\columnwidth]{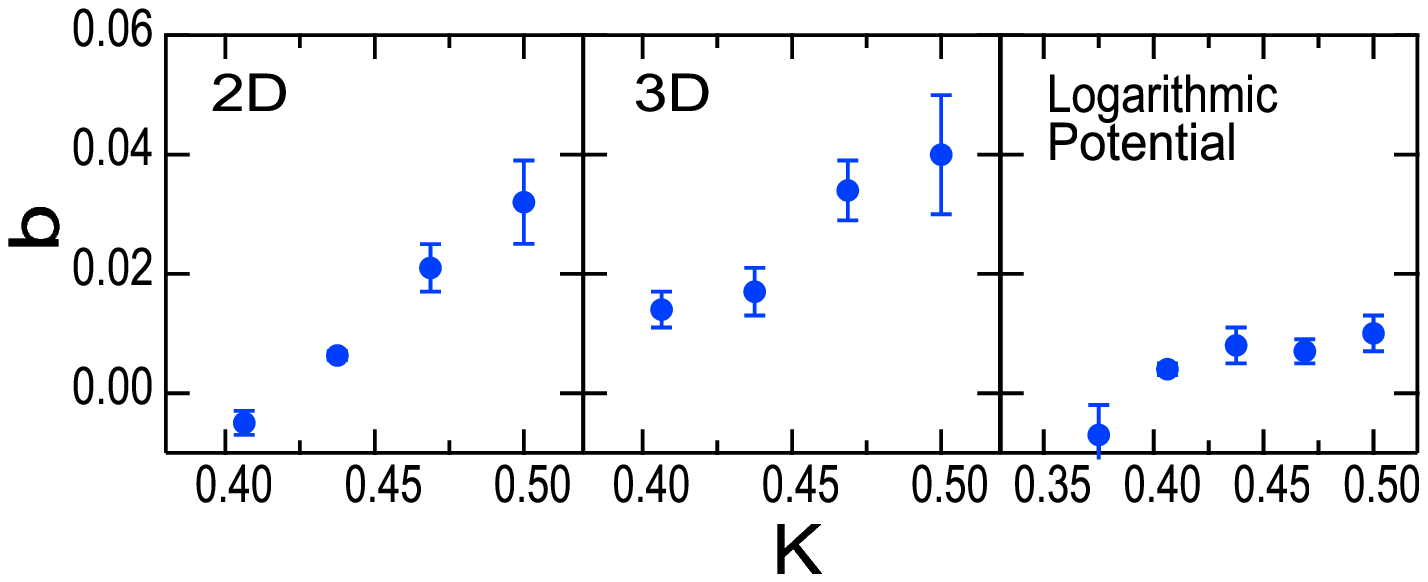} \vskip 0.1cm
(b) \includegraphics[width=0.93\columnwidth]{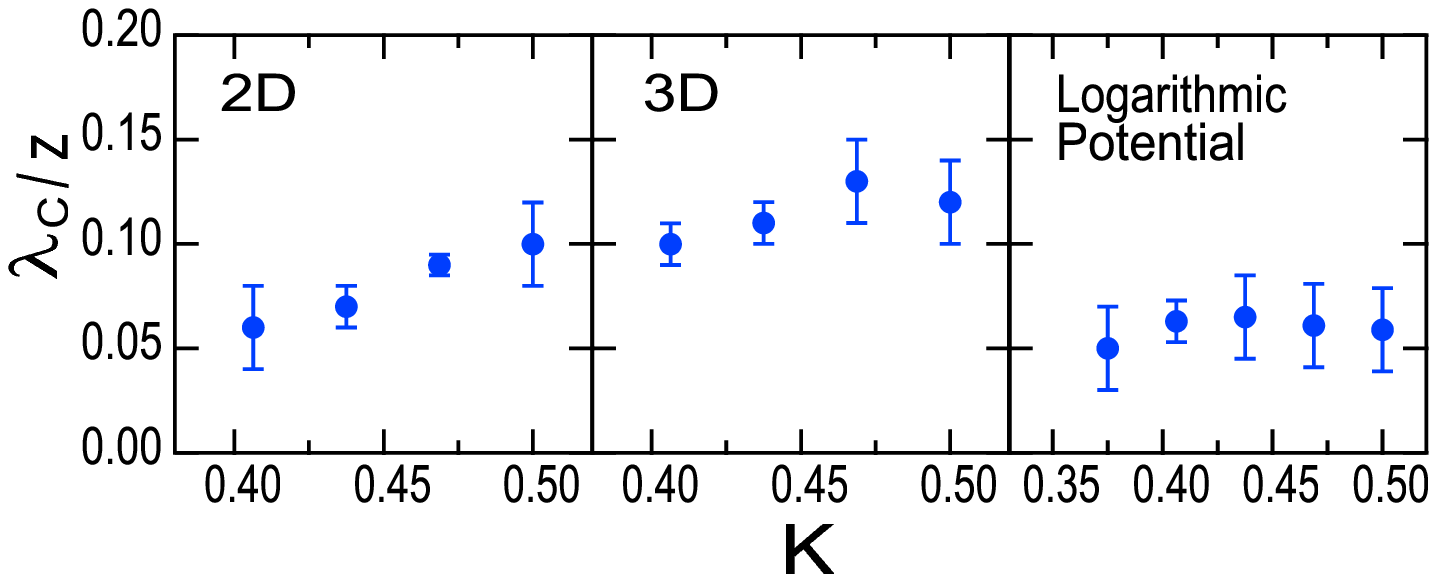} \vskip 0.1cm
(c) \includegraphics[width=0.93\columnwidth]{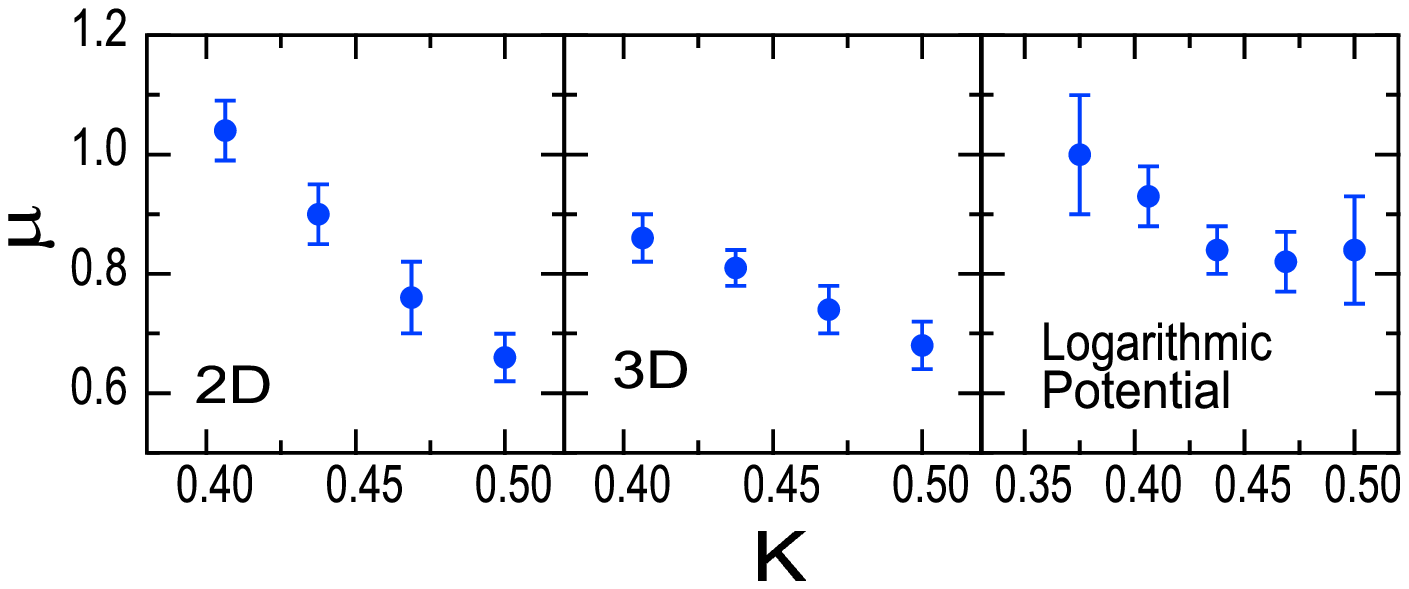}
\caption{Aging scaling exponents at $T = 0.02$ as functions of $K$ for the
   Coulomb glass ($1/r$ interaction) in two (left panels) and three dimensions
   (center), and for the Bose glass (with logarithmic repulsion, right):
   (a) full aging scaling exponent $b$, Eq.~(\ref{gags}) with $\mu = 1$;
   (b) autocorrelation decay exponent $\lambda_C / z$, Eq.~(\ref{scfd});
   (c) subaging exponent $\mu$, Eq.~(\ref{gags}) with $b = 0$.} 
\label{fig9}
\end{figure}
Intriguingly, our data reveal that the aging scaling exponents also
depend on the total charge carrier density $K$.
As evidenced in Fig.~\ref{fig9}, the non-equilibrium relaxation processes from 
the initial high-temperature configurations proceed increasingly slower as the
filling fraction $K$ is tuned away from $K = 1/2$. 
These plots list our results measured at $T = 0.02$ for $K = 0.40625$, 
$0.4375$, $0.46875$, and $K = 0.5$ for the Coulomb glass model in two (left 
panels) and three dimensions (center panels), as well as in addition for 
$K = 0.375$ for the two-dimensional Bose glass system; but recall that owing 
to particle-hole symmetry the same data apply for both $K = 0.5 \pm k$ above 
and below half-filling.
At the lowest filling fractions we investigated for the two-dimensional 
Coulomb and Bose glasses, we already obtain unphysical values $b < 0$ and
correspondingly $\mu > 1$: 
These systems at $K = 0.40625$ and $0.375$, respectively, are already frozen
in on the time domain accessible to our Monte Carlo simulations.
We are hence limited to the carrier density range $0.4 < K < 0.6$.

Within the full-aging scaling analysis, Figs.~\ref{fig9}(a) and (b), it is
apparent that the Bose glass exponents display a much weaker dependence on 
the filling fraction than is visible for either the two- or three-dimensional
Coulomb glass.
We tentatively attribute this observation to the considerably wider soft gap 
in the density of states that emerges for the logarithmic interaction 
potential as compared with the Coulomb $1/r$ repulsion, compare 
Figs.~\ref{fig2}(a) and \ref{fig4}(a).
In the long-time aging scaling regime, spatial rearrangements only redistribute
energy levels deep inside this Coulomb gap, which attains a much more
$K$-independent shape and still remains very shallow for the Bose glass in, 
e.g., the interval $|\epsilon - \mu_c| \leq 0.5$, for which the effects of
modified filling fractions already become clearly discernible in the Coulomb
glass.
Remarkably, though, our data yield a noticeable dependence of the subaging
scaling exponent $\mu$ even for the Bose glass with logarithmic interactions.

Consequently, the aging scaling exponents in the Coulomb and Bose glass appear
to be non-universal, depending both on temperature and filling fraction, aside
from dimensionality and the form of the long-range repulsive potential.
Non-universal aging scaling has also been observed in other disordered systems,
as for example the two-dimensional random-site \cite{Park10} and random-bond
\cite{Henkel08, Park12} Ising models or the three-dimensional Edwards-Anderson
spin glass with a bimodal distribution of the coupling constants 
\cite{Kisker96, Park12}, where some of the scaling exponents were found to 
depend on temperature and/or the disorder. 
Our present work therefore provides additional interesting examples of 
disordered systems that display non-universal aging exponents.

\section{Summary and Conclusions}
\label{concl}

We have carefully investigated non-equilibrium relaxation processes and aging
scaling of the Coulomb glass model in two and three dimensions, and of the
Bose glass system in two dimensions through Monte Carlo simulations at low
temperatures.
We confirm that the long-time dynamics in the $\alpha$ relaxation regime for
the two-time autocorrelation function can be described by the simple general
aging scaling form (\ref{gags}). 
We have employed either full-aging or subaging simplified scaling forms, and
assess that neither version appears to provide substantially superior scaling
collapse, although on physical grounds we tend to prefer full-aging scaling
described by Eq.~(\ref{gags}) with $\mu = 1$ and Eq.~(\ref{scfd}).
The extracted aging scaling exponents depend on the filling fraction and
temperature, in addition to dimensionality and form of the repulsive 
interaction potential, and are hence not universal.
Moreover they follow a common trend: 
We observe that as either the temperature decreases or the charge carrier 
density deviates more from half-filling, the aging exponents reflect 
considerably slowed-down relaxation kinetics.

A series of recent studies \cite{Park10, Park12, Corberi12} has shown that in
disordered coarsening systems governed by a single length scale $L(t)$ one
typically encounters rather complicated growth laws, characterized by a 
cross-over from a transient power-law growth to asymptotically logarithmic 
growth. 
Using this length $L(t)$ as variable in the aging scaling analysis reveals 
that the full aging scenario prevails in these systems. 
It it an interesting question whether a similar cross-over between different 
growth regimes also exists in the Coulomb and Bose glasses, for which we
would tentatively interpret $L(t)$ to describe the emerging spatial
(anti-)correlations as the mutually repelling particles relax towards more
energetically favorable sites.
One way to extract a time-dependent length is through an analysis of the 
space-time correlation function. 
Computing this correlation function is a challenging task for our off-lattice 
model with long-range repulsive interactions. 
Because of the importance of this length in the non-equilibrium relaxation 
process, we plan to come back to this issue in the future.

\acknowledgments
This research is supported by the U.S. Department of Energy, Office of Basic 
Energy Sciences, Division of Materials Sciences and Engineering under Award 
DE-FG02-09ER46613.

\end{document}